\documentclass[fleqn,usenatbib]{style/mnras}

\usepackage{mathptmx}
\usepackage{txfonts}
\usepackage{type1ec}
\usepackage[pdftex]{graphicx}
\usepackage[T1]{fontenc}
\usepackage{ae,aecompl}

\DeclareRobustCommand{\VAN}[3]{#2}
\let\VANthebibliography\thebibliography
\def\thebibliography{\DeclareRobustCommand{\VAN}[3]{##3}\VANthebibliography}

\usepackage{amssymb}	% Extra maths symbols
\usepackage[usenames,dvipsnames]{xcolor}
\usepackage{mathrsfs}

\newcommand{\mdota}{\dot{M}_{\rm Acc}}
\newcommand{\mdotw}{\dot{M}_{\rm Wind}}
\newcommand{\msun}{\rm M_{\rm \sun}}
\newcommand{\vk}{{\rm v}_{\rm K}}
\newcommand{\Elsasser}{\Lambda}
\newcommand{\ElsasserO}{\Lambda_{\rm O}}
\newcommand{\ElsasserH}{\Lambda_{\rm H}}
\newcommand{\ElsasserA}{\Lambda_{\rm A}}
\newcommand{\ElsasserP}{\Lambda_{\rm P}}
\newcommand{\etaA}{\eta_{\rm A}}
\newcommand{\etaH}{\eta_{\rm H}}
\newcommand{\etaO}{\eta_{\rm O}}
\newcommand{\amin}{a_{\rm min}}
\newcommand{\amax}{a_{\rm max}}
\newcommand{\hcop}{\rm HCO^{\rm +}}
\newcommand{\ntwohp}{\rm N_{\rm 2}H^{\rm +}}

\title[Formation of rings with disc chemistry]{On the importance of disc chemistry in the formation of protoplanetary disc rings}

\author[C. A. Nolan et al.]{
C. A. Nolan$^{1}$\thanks{E-mail: chrisnolan.au@gmail.com},
B. Zhao$^{1,2}$,
P. Caselli$^{1}$,
Z. Y. Li$^{3}$
\\
$^{1}$Max-Planck-Institut f\"{u}r extraterrestrische Physik (MPE), Giessenbachstr 1, D-85748 Garching, Germany \\
$^{2}$Department of Physics \& Astronomy, McMaster University, Hamilton, ON L8S 4K1, Canada\\
$^{3}$Department of Astronomy, University of Virginia, 530 McCormick Road, Charlottesville, VA 22904, USA
}

\date{Accepted XXX. Received YYY; in original form ZZZ}

\pubyear{2023}

\begin{document}
\label{firstpage}
\pagerange{\pageref{firstpage}--\pageref{lastpage}}
\maketitle{}

\begin{abstract}
Radial substructures have now been observed in a wide range of protoplanetary discs (PPDs), from young to old systems, however their formation is still an area of vigorous debate. Recent magnetohydrodynamic (MHD) simulations have shown that rings and gaps can form naturally in PPDs when non-ideal MHD effects are included. However these simulations employ ad-hoc approximations to the magnitudes of the magnetic diffusivities in order to facilitate ring growth. We replace the parametrisation of these terms with a simple chemical network and grain distribution model to calculate the non-ideal effects in a more self-consistent way. We use a range of grain distributions to simulate grain formation for different disc conditions. Including ambipolar diffusion, we find that large grain populations (> 1$\mu$m), and those including a population of very small polyaromatic hydrocarbons (PAHs) facilitate the growth of periodic, stable rings, while intermediate sized grains suppress ring formation. Including Ohmic diffusion removes the positive influence of PAHs, with only large grain populations still producing periodic ring and gap structures. These results relate closely to the degree of coupling between the magnetic field and the neutral disc material, quantified by the non-dimensional Elsasser number $\Elsasser$ (the ratio of magnetic forces to Coriolis force). For both the ambipolar-only and ambipolar-ohmic cases, if the total Elsasser number is initially of order unity along the disc mid-plane, ring and gap structures may develop. 
\end{abstract}


%%%%%%%%%%%%%%%%%%%%%%%%%%%%%%%%%%%%%%%%%%%%%%%%%%

%%%%%%%%%%%%%%%%% BODY OF PAPER %%%%%%%%%%%%%%%%%%

%=================================================
% INTRODUCTION 
%=================================================
\section{Introduction}

% The ubiquity of rings in PPDs and the various different proposed formation mechanisms for these rings
Recent observations with the Atacama Large Millimeter/submillimeter Array (ALMA) have revealed that most, if not all protoplanetary discs are highly structured \citep{ALMAPartnershipEtAl2015, HuangEtAl2018, Andrews2020}. These structures include rings and gaps \citep[e.g.][]{AndrewsEtAl2016, LongEtAl2018, AndrewsEtAl2018, HuangEtAl2018, AvenhausEtAl2018, VillenaveEtAl2019, PerezEtAl2020}, central cavities \citep[e.g.][]{CasassusEtAl2013, PinillaEtAl2017, PinillaEtAl2018, vanderMarelEtAl2018, KudoEtAl2018, FacchiniEtAl2020}, spirals \citep[e.g.][]{HashimotoEtAl2011, GarufiEtAl2013, PerezEtAl2016, StolkerEtAl2017, HuangEtAl2018a}, and azimuthal asymmetries/arcs \citep[e.g.][]{vanderMarelEtAl2013, PerezEtAl2014, KrausEtAl2017, DongEtAl2018, CazzolettiEtAl2018, IsellaEtAl2018}. While interactions between forming planets and the disc is a prominent explanation for the existence of these structures \citep[e.g.][]{DongEtAl2015, DipierroEtAl2015a, BaeEtAl2017, DongEtAl2017}, many alternative mechanisms have been proposed and produce comparable structures. For rings and gaps, which are the most commonly detected \citep{PinteEtAl2022}, these mechanisms include condensation fronts \citep{ZhangEtAl2015}, dust sintering \citep{OkuzumiEtAl2016}, thermal wave instabilities \citep{UedaEtAl2021}, baroclinic instabilities \citep{KlahrBodenheimer2003}, the secular gravitational instability \citep{TakahashiInutsuka2016}, counter-rotating infall \citep{VorobyovEtAl2016}, and zonal flows \citep{BethuneEtAl2017}.
% See SurianoEtAl2018, PinteEtAl2022, BenistyEtAl2022, HuEtAl2022 for refs

% Ring formation is a natural consequence of non-ideal MHD discs
If magnetized winds are the primary angular momentum transport mechanism in non-ideal regions of PPDs, as has been the dominant train of thought for the last decade \citep[e.g.][]{LesurEtAl2022}, it is highly likely that the disc spontaneously organizes into regularly spaced rings and gaps \citep{BethuneEtAl2017, SurianoEtAl2019, RiolsEtAl2020, CuiBai2021}. In such discs, an initial small perturbation of gas density leads to the radial advection of vertical magnetic flux. This leads to a local concentration of flux in the less dense regions.
%due to viscous diffusion. 
If these higher flux/lower density regions drive stronger mass loss and faster accretion due to more efficient removal of angular momentum, the density further decreases, reinforcing the initial density perturbation \citep{LubowEtAl1994, RiolsLesur2019}. This results in MHD wind-driven spontaneous ring and gap formation in discs.

% Justification: adding chemistry to see if rings still form
While the discovery of spontaneous ring and gap formation as a natural consequence of MHD wind-driven discs is a significant result, it is important to explore the conditions that are required to trigger this mechanism, in order to confirm that it is applicable to realistic PPDs, and for what conditions in these discs it can operate. In a series of papers, \citet{SurianoEtAl2017, SurianoEtAl2018, SurianoEtAl2019} explored the viability of wind-driven ring formation for different values of field-neutral coupling $\Elsasser$, plasma-$\beta$, and the inclusion of Ohmic diffusion. They find that for initial inner disc values of between $0.05 < \Elsasser < 0.5$, prominent rings and gaps are formed within the disc. While they find little evidence for ring formation at $\Elsasser \leq 0.01$, for $\Elsasser > 0.5$ and in the ideal MHD limit they note that ring/gap formation is present with unsteady disc accretion and outflows, highlighting the robustness of wind-driven ring formation. \citet{SurianoEtAl2019} showed that spontaneous long-term ring formation still occurs in 3D simulations, however spiral structures begin to dominate at larger $\Elsasser$ (inner disc $\Elsasser \geq 1.25$). \citet{BethuneEtAl2017} also generated stable rings/gaps in 3D simulations using a small chemical network to calculate the ionization fraction. Interestingly, with the inclusion of Hall diffusion they find that ring formation is inhibited for reversed vertical magnetic field configurations (i.e. $\mathbf{\Omega} \cdot \mathbf{B} < 0$). 

In this study, we focus on the calculation of more accurate diffusion coefficients, adopting a simplified chemical network and grain distribution model to replace the parametrised diffusivities used in previous studies \citep{SurianoEtAl2017, SurianoEtAl2018, SurianoEtAl2019, RiolsEtAl2020, CuiBai2021, HuEtAl2022}, to see how grain distribution affects the formation of rings in discs. We find that rings only form for neutral-field coupling values of $\Elsasser \gtrsim 1$, corresponding to more evolved grain populations with minimum sizes of 1 $\mu$m. This has important implications for the classes of discs in which this instability may be triggered.
% Papers to check for important results: BethuneEtAl2017,SurianoEtAl2017,2018,2019,RiolsEtAl2020,CuiBai2021 

% Paper layout
This paper is organized as follows. In Section \ref{sec:method}, we describe the simulation setup, including the MHD equations, disc initial conditions, simulation grid and boundary conditions. In Section \ref{sec:chem} we describe the chemical model used to calculate the magnetic diffusion terms, including the ionization model and chemical network. In Section \ref{sec:diffStruct} we present the different grain distributions tested in the paper and the resulting initial ambipolar diffusion disc profiles, while in Section \ref{sec:discEvol} we summarize the results of evolving these simulations in time. In Section \ref{sec:ohmicbeta} we explore how the addition of Ohmic diffusion and changing the plasma-$\beta$ modify the results from Section \ref{sec:discEvol}, while in Section \ref{sec:discussion} we summarise the results and compare them with other numerical works and observations, proposing a robust requirement for the formation of rings in non-ideal discs. Finally, Section \ref{sec:conclusions} concludes with the main results of the study.

%=================================================
% PROBLEM SETUP 
%=================================================
\section{Problem Setup} \label{sec:method}

%-------------------------------------------------
% MHD equations
%-------------------------------------------------
\subsection{MHD Equations}

We use the \textsc{zeustw} code \citep{KrasnopolskyEtAl2010} to solve the time-dependent, non-ideal MHD equations in axisymmetric spherical coordinates ($r$, $\theta$, $\phi$). These equations describe the conservation of mass
\begin{equation}
\frac{\partial \rho}{\partial t} + \nabla \cdot (\rho \mathbf{v}) = 0, \label{eqn:masscons}
\end{equation}
and the conservation of momentum
\begin{equation}
\rho \frac{\partial \mathbf{v}}{\partial t} + \rho (\mathbf{v} \cdot \nabla) \mathbf{v} = -\nabla P + \mathbf{J} \times \mathbf {B}/c - \rho \nabla \Phi_{\rm g},
\end{equation}
for the neutral gas, as well as the evolution of the magnetic field $\mathbf{B}$ via the induction equation
\begin{equation}
\frac{\partial \mathbf{B}}{\partial t} = \nabla \times (\mathbf{v} \times \mathbf{B}) - \frac{4 \pi}{c} \nabla \times (\eta_{O} \mathbf{J} + \eta_{A} \mathbf{J}_{\perp}), \label{eqn:induction}
\end{equation}
and lastly the conservation of energy
\begin{equation}
\frac{\partial e}{\partial t} + \nabla \cdot (e \mathbf{v}) = - P \nabla \cdot \mathbf{v}. \label{eqn:energycons} 
\end{equation}
In the above equations, $\rho$ is the density, $\mathbf{v}$ is the velocity, and the internal energy $e = P/(\Gamma - 1)$, where $P$ is the thermal pressure and $\Gamma$ is the adiabatic index. The current density is $\mathbf{J} = (c/4\pi)\nabla \times \mathbf{B}$, and the component of the current density perpendicular to the magnetic field $\mathbf{J}_{\perp} = -(\mathbf{J} \times \mathbf{B}) \times \mathbf{B}/B^2$. In addition, $\Phi_{\rm g}$ is the gravitational potential of the central object, given by $\Phi_{\rm g} = -G M/r$, where $G$ is the gravitational constant and $M$ is the mass of the protostar (which is treated as a point mass at the origin of the coordinate system). Finally, $\eta_{O}$ is the Ohmic resistivity and $\eta_{A}$ is the ambipolar diffusivity. In the \textsc{zeustw} code, Ohmic resistivity is applied using the algorithm of \citet{FlemingEtAl2000}, while ambipolar diffusion is applied using the fully explicit method of \citet[see also \citealt{LiEtAl2011}]{MacLowEtAl1995}. Sub-cycling was employed to speed up the treatment of ambipolar diffusion. Sometimes the ambipolar diffusion time step becomes prohibitively small in the polar region because of an unusually large ambipolar diffusivity caused by a low density and strong magnetic field. In such cases, we cap the ambipolar diffusivity through a minimum AD time step, as done in \cite{ZhaoETAL2021}. The cap has little effect on the much denser disk, the focus of our investigation.

%-------------------------------------------------
% Initial Conditions
%-------------------------------------------------
\subsection{Initial Conditions}

The initial conditions for all simulations are similar to those described in \citet{SurianoEtAl2018}. For completeness, we describe them here in modest detail.

The simulation domain is divided into two regions: a thin, cold, rotating disc orbiting a 1 $M_{\sun}$ protostar, and an initially non-rotating, hot corona above the disc, which is rapidly replaced by a magnetic wind driven out from the disc. To facilitate comparison with similar studies of substructure formation in non-ideal MHD disks in the literature \citep[e.g.][]{SurianoEtAl2018,RiolsEtAl2020,HuEtAl2022}, we choose $\Gamma = 1.01$ to approximate the oft-adopted local isothermal conditions, i.e., that each initial parcel of material retains most of its initial temperature independent of its future location; more detailed temperature calculations are needed to quantify the effects of this approximation on the substructure formation. We assume that the initial temperature distribution decreases with radius according to $T \propto r^{-1}$, so that the sound speed is proportional to the local Keplerian speed.

% Disc -------------------------------------------
\subsubsection{Disc}

We characterize the disc by the dimensionless parameter $\epsilon = h/r = c_{\rm s}/\vk = 0.05$, where $h$ is the disc scale height, $c_{\rm s}$ is the isothermal sound speed, and $\vk$ is the Keplerian velocity. The disc region is defined within $\theta \in [\pi/2 - \theta_0, \pi/2 + \theta_0]$, with the disc half opening angle set to $\theta_0 = {\rm arctan}(2 \epsilon)$, i.e. two scale heights. Assuming hydrostatic balance, the disc density then takes on the form
\begin{equation}
\rho_{\rm d}(r, \theta) = \rho_0\left(\frac{r}{r_0}\right)^{-\alpha_{\rm d}} \exp \left(-\frac{\cos^2\theta}{2 \epsilon^2} \right),
\end{equation}
where the exponential term can be simplified to a standard Gaussian profile in the cylindrical coordinate $z$. The subscript `0' denotes mid-plane values at the inner radial boundary, and for all simulations we assume $\alpha_{\rm d} = 3/2$. The disc pressure is defined by 
\begin{equation}
P_{\rm d}(r, \theta) = \rho_{\rm d}(r,\theta) c_{\rm s}^2,
\end{equation}
where $c_{\rm s} = \epsilon \vk$. As a result of the radial pressure gradient, and in order to maintain hydrostatic equilibrium, the initial azimuthal velocity within the disc is set to be slightly sub-Keplerian:
\begin{equation}
{\rm v}_{\phi} = \vk \sqrt{1 - \epsilon^2(1+\alpha_{\rm d})}.
\end{equation}

% Corona -----------------------------------------
\subsubsection{Coronal region}

The coronal region, defined above and below the disc region, is required to maintain hydrostatic equilibrium with the disc. Therefore we set the coronal density and pressure as 
\begin{equation}
\rho_{c}(r) = \rho_0 \epsilon^2 (1+\alpha_{\rm d}) \exp \left[-\frac{\cos^2\theta_0}{2 \epsilon^2} \right] \left(\frac{r}{r_0}\right)^{-\alpha_{\rm d}} \equiv \rho_{c,0}\left(\frac{r}{r_0}\right)^{-\alpha_{\rm d}},
\end{equation}
and
\begin{equation}
P_{c}(r) = \rho_{c}(r) \vk^2/(1 + \alpha_{\rm d}),
\end{equation}
respectively, in order to maintain the pressure balance at the disc surface. It is worth noting that while the hot coronal material is important for the initial setup, the coronal gas is rapidly replaced by colder material from the disc, and does not effect the long-term evolution of the simulations.

% Magnetic field ---------------------------------
\subsubsection{Magnetic field} \label{sec:magfield}

To ensure that the magnetic field is divergence-free initially, we set the poloidal magnetic field components using the magnetic flux function $\Psi$ as in \citet{ZanniEtAl2007}, 
\begin{equation}
\Psi(r, \theta) = \frac{4}{3} r_0^2 B_{\rm p, 0} \left( \frac{r \sin \theta}{r_0} \right)^{3/4} \frac{m^{5/4}}{(m^2 + \cot^2 \theta)^{5/8}},
\end{equation}
where $B_{\rm p,0}$ sets the magnetic field at $(r_0, \pi/2)$, and the parameter $m$ determines the vertical scale on which the initial magnetic field bends, where a value $m \rightarrow \infty$ gives a perfectly vertical field. We indirectly set the value of $B_{\rm p,0}$ by assigning the initial plasma-$\beta$, the ratio of the thermal to magnetic pressure at the disc mid-plane, which is set to $0.922 \times 10^3$ for most simulations, in keeping with \citet{SurianoEtAl2018}. We also use $m = 0.5$ for all simulations presented in this paper. The initial magnetic field components are then calculated by the following equations:
\begin{eqnarray}
B_r = \frac{1}{r^2 \sin \theta} \frac{\partial \Psi}{\partial \theta}, \\
B_{\theta} = - \frac{1}{r \sin \theta}\frac{\partial \Psi}{\partial r}.
\end{eqnarray}

%-------------------------------------------------
% Simplified Ambipolar Diffusion
%-------------------------------------------------
\subsection{Simplified Ambipolar Diffusion} \label{sec:simpAD}

In order to compare our results with the reference simulation ad-els0.25 presented in \citet{SurianoEtAl2018}, we use an identical simulation setup except for the calculation of the diffusivities. \citet{SurianoEtAl2018} use a parametrized formulation for the density of ions, in order to fix the initial dimensionless ambipolar Elsasser number
\begin{equation}
\ElsasserA = \frac{{\rm v}_{\rm A}^2}{\Omega_{\rm K} \etaA}, \label{eqn:lambdaa}
\end{equation}
where ${\rm v}_{\rm A}$ is the Alfv\'{e}n velocity. Physically, the Elsasser number is the ratio of the collision frequency of ions and neutral particles to the natural frequency scale in protoplanetary discs, the Keplerian orbital frequency $\Omega_{\rm K}$, and measures the degree of coupling between the magnetic field and the neutral disc material. For values $\gg 1$, the magnetic field and hence the ions are essentially frozen into the neutral material, while for values $\ll 1$, the neutrals are effectively decoupled from the magnetic field. 
In their reference simulation ad-els0.25, \citet{SurianoEtAl2018} assume that $\ElsasserA = 0.25$ at the inner boundary on the disc mid-plane, and set $\ElsasserA$ proportional to $r^{3/4}$, i.e. the disc material at larger radii is better coupled to the magnetic field than that at smaller radii. \citet{SurianoEtAl2018} also include an explicit isotropic Ohmic resistivity $\etaO$ in addition to ambipolar diffusion in a number of their simulations. Of interest to us is the simulation oh2.6, which we use to compare our results to in Section \ref{sec:ohmicbeta}. In this simulation the ohmic diffusion is constant in time and space and has a value of $\etaO = 2.5 \times 10^{15}$ cm$^2$ s$^{-1}$.

%-------------------------------------------------
% Simulation grid
%-------------------------------------------------
\subsection{Simulation grid}

Equations (\ref{eqn:masscons}) - (\ref{eqn:energycons}) are solved for $r \in\left[1,100\right]$ au and $\theta \in \left[0,\pi\right]$. Given the focus on ambipolar diffusion, the radial grid extents are chosen such that they cover the regions of the disc where AD is anticipated to be the dominant non-ideal effect \citep{TurnerEtAl2014}. We use a geometrically spaced grid in the radial direction such that ${\rm d}r_{i+1}/{\rm d}r_i = 1.012$ is constant and $r_{i+1} = r_i + {\rm d}r$, with the grid spacing at the inner edge set as ${\rm d}r_0 = 2.3 r_0 {\rm d}\theta$. The grid is uniform in $\theta$, with a resolution of ${\rm n}_r$ $\times$ ${\rm n}_{\theta} = 400$ $\times$ $720$, the same as \citet{SurianoEtAl2018}. This results in 12 grid cells per vertical scale height, and 48 cells between lower and upper disc surfaces.

%-------------------------------------------------
% Boundary Conditions
%-------------------------------------------------
\subsection{Boundary Conditions} \label{sec:bound}

In keeping with the simulations of \citet{SurianoEtAl2018}, we use standard outflow conditions for the inner and outer radial domain boundaries. In this configuration, the flow quantities in the first active zone within the simulation domain are copied into the ghost zones\footnote{Ghost zones are the dummy cells outside the simulation domain that are assigned values to create the required conditions at the simulation boundary}, except for the radial component of the velocity, ${\rm v}_r$, which is set to zero in the ghost zones if it points into the computational domain in the first active zone. For the axial boundaries ($\theta = 0$ and $\pi$), we use standard axial reflection conditions, where the density and radial components of the velocity and magnetic field (${\rm v}_r$ and $B_r$) in the ghost zones take their values in the corresponding active zones, while the polar and azimuthal components (${\rm v}_{\theta}$, $B_{\theta}$, and ${\rm v}_{\phi}$) take the negative of their values in the corresponding active zones\footnote{Ideally, one should impose polar boundary conditions on the magnetic field through the emf $\mathbf{\epsilon}\equiv \mathbf{v}\times \mathbf{B}$, as done in, e.g., \cite{KrasnopolskyEtAl1999} to preserve the divergence-free condition to the machine accuracy. However, we find through experimentation that the simpler axial reflection boundary conditions do not lead to numerical problems associated with potential magnetic monopoles.}. 
The axial reflective boundary condition on the magnetic field guarantees that no poloidal field lines leave the simulation domain through the $\theta$ boundaries. Portions of some field lines can exit the inner radial boundary, but this does not reduce the overall poloidal magnetic flux since the rest of such field lines still thread the simulation domain. We set $B_{\phi}$ to vanish on the polar axis and the inner radial boundary, since it is taken to be non-rotating.

We also note here that in order to keep the duration of the simulation to a manageable time span we restrict the minimum density allowed for any given cell to 
\begin{equation}
\rho_{\rm min} = \frac{B^2}{4 \pi \left(|\Delta x|_{\rm min}/{\rm d}t_{\rm floor, Alfv\acute{e}n} \right)^2},
\end{equation}
where $|\Delta x|_{\rm min}$ is the smallest of the cell's dimensions along $r$ and $\theta$, and the minimum time step for any cell, ${\rm d}t_{\rm floor, Alfv\acute{e}n}$ is set to $1 \times 10^4$ s. As a result, artificial mass is added to cells with densities below the minimum density threshold. This density floor is usually triggered in the central regions ($r < 10$ au) with very large Alfv\'{e}n speeds. 

%=================================================
% CHEMISTRY MODEL
%=================================================
\section{Chemistry Model} \label{sec:chem}

The magnetic diffusivities within protoplanetary discs are determined by the chemistry and microphysical processes, including external and internal ionization processes and thermal collisions between the different species. While simple prescriptions for the diffusivities, such as those listed in Section \ref{sec:simpAD}, can give a first-order approximation to the properties of protoplanetary discs and winds, chemical networks can much more accurately determine the ionization fraction and hence the relative importance of the different diffusion coefficients. To this effect we adopt a simple equilibrium chemical network based on \citet{ZhaoEtAl2018a}, which is described below.

%-------------------------------------------------
% Ionisation rate
%-------------------------------------------------
\subsection{Ionisation rate}

%Intro
We model the ionization field based on three of the main sources of ionization in PPD's: X-rays, cosmic rays, and radioactivity from within the disc itself. In the simulations, we trace radial rays from the central star on the spherical grid to obtain the column densities crossed by the rays, denoted by $\Sigma_{r}(r,\theta)$. Similarly, we trace $\theta$-rays from the upper and lower poles towards the disc at constant-$r$, and define two column densities $\Sigma_{\theta}^{\rm top}(r, \theta)$ and $\Sigma_{\theta}^{\rm bot}(r, \theta)$. While these rays are not straight, given the assumption of a geometrically thin disc considered here, the two column densities only reach physically meaningful values where the rays are largely vertical, i.e. the region near the disc mid-plane.

%X-rays
For X-ray ionization, we use the spherical formulation \citep{Bai2017} of the fitted formula of \citet{BaiGoodman2009}, based on the calculations of \citet{IgeaGlassgold1999}. We use the fitting coefficients at an X-ray temperature of $T_{\rm X} = 3$ keV, giving
\begin{eqnarray}
\zeta_{\rm X} = \left( \frac{r}{1 \mbox{ au}} \right)^{-2.2} \frac{L_{\rm X}}{10^{30} \mbox{ erg s}^{-1}} \lbrace \zeta_1 e^{-(\Sigma_r/5\Sigma_{X,a})^{\alpha}} \nonumber \\
+ \zeta_2 \lbrack e^{-(\Sigma^{\rm top}_{\theta}/\Sigma_{X,s})^{\beta}} + e^{-(\Sigma^{\rm bot}_{\theta}/\Sigma_{X,s})^{\beta}} \rbrack \rbrace \mbox{ s}^{-1}. \label{eqn:xrays}
\end{eqnarray}
The first term in equation (\ref{eqn:xrays}) describes direct absorption of X-rays along the radial direction, with $\zeta_1 = 6.0 \times 10^{-11}$ s$^{-1}$, $\Sigma_{X,a} = 3.6 \times 10^{-3}$ g cm$^{-2}$, and $\alpha = 0.4$. The factor of 5 originates from the fact that the $\Sigma_{X,a}$ value in the original fitting formula corresponds to the vertical instead of the radial column density; the factor of 5 accounts for the conversion between the two. The second term accounts for scattered X-rays, with $\zeta_2 = 1.0 \times 10^{-14}$ s$^{-1}$, $\Sigma_{X,s} = 1.7$ g cm$^{-2}$, $\beta = 0.65$. For all models we adopt $L_{X} = 2.34 \times 10^{30}$ ergs s$^{-1}$ as the X-ray luminosity expected for a 1~$M_{\sun}$ star according to the relation found by \citet{PreibischEtAl2005}:
\begin{equation}
\log_{10} \left(L_{X} \lbrack \mbox{erg s}^{-1}\rbrack \right) = 30.37 + 1.44 \times \log_{10} \left(M/M_{\sun} \right),
\end{equation}
which was derived empirically from T-Tauri sources in the Orion Nebula Cluster.

%Cosmic rays
For the calculation of the cosmic-ray ionization, we use the recent model of \citet{PadovaniEtAl2018}, which takes into account the propagation of primary and secondary cosmic-ray particles. It is given by the following fitting formula:
\begin{equation}
\frac{\zeta_{\rm CR}}{\rm s^{-1}} = 10^{\sum_{k \ge 0} c_k  \left(\log_{10} \frac{\Sigma_{\theta}^{\rm top}}{\delta}\right)^{k}} + 10^{\sum_{k \ge 0} c_k  \left(\log_{10} \frac{\Sigma_{\theta}^{\rm bot}}{\delta}\right)^{k}}, 
\end{equation}
where $\delta = 3.95 \times 10^{-24}$ g cm$^{-2}$. Table \ref{tab:cr} gives the set of coefficients $c_k$ for the $\mathscr{L}$ model, which approximates the cosmic-ray proton spectrum by extrapolating the Voyager 1 data \citep[see][]{IvlevEtAl2015,PadovaniEtAl2018}, and is considered a lower bound to the actual average Galactic cosmic-ray spectrum. \citet{PadovaniEtAl2018} demonstrated that for $\Sigma \lesssim 130$ g cm$^{-2}$, the effective column density is not line-of-sight, but follows the magnetic field. If the magnetic field lines are strongly twisted, this can lead to effective column densities much larger than the line-of-sight column density at that point \citep[e.g.][]{PadovaniGalli2011,PadovaniGalli2013,PadovaniEtAl2013}. For simplicity, we assume that the surface density for all magnitudes is line-of-sight. We also set a maximum cosmic-ray ionization rate of $\zeta_{\rm CR} = 3 \times 10^{-17}$ s$^{-1}$ to prevent very large ionization rates near the polar axis, where $\Sigma_{\theta} \to 0$. This gives an effective minimum column density of $\Sigma_{\rm min} = 0.01$ g cm$^{-2}$.

\begin{table}
\caption[]{Coefficients $c_k$ for the $\mathscr{L}$ cosmic-ray model of \citet{PadovaniEtAl2018}.}
\begin{center}
\begin{tabular}{c c}
\hline
\hline
k& model $\mathscr{L}$\\
\hline
0 & $-3.331056497233 \times 10^6$\\
1 & $ 1.207744586503 \times 10^6$ \\
2 & $-1.913914106234 \times 10^5$ \\
3 & $ 1.731822350618 \times 10^4$ \\
4 & $-9.790557206178 \times 10^2$ \\
5 & $ 3.543830893824 \times 10^1$ \\
6 & $-8.034869454520 \times 10^{-1}$ \\
7 & $ 1.048808593086 \times 10^{-2}$ \\
8 & $-6.188760100997 \times 10^{-5}$ \\
9 & $ 3.122820990797 \times 10^{-8}$ \\
\hline
\end{tabular}
\end{center}
\label{tab:cr}
\end{table}

% Radioactive nuclides
Finally, we added an ionization component due to radioactive nuclides within the disc according to
\begin{equation}
\zeta_{\rm R} = 1.1 \times 10^{-22} \mbox{ s}^{-1},
\end{equation}
based on the ionization rate of the main source of ionization among the long-lived radionuclides, $^{40}$K \citep[half-life $1.3 \times 10^9$ yr][]{UmebayashiNakano2009}. The main source of ionization among the short-lived radioactive nuclides, $^{26}$Al, has an ionization rate $\sim 10^3$ times that of $^{40}$K, however its half-life is comparatively short: only $7.4 \times 10^5$ yrs. Since we are modelling discs out of the embedded phase (Class I and older), we expect that the ionizing effect of the short-lived radionuclides will be substantially diminished, and hence why we exclude them from this model. The total ionization rate is then simply $\zeta = \zeta_{\rm X} + \zeta_{\rm CR} + \zeta_{\rm R}$.

%-------------------------------------------------
% Chemical network
%-------------------------------------------------
\subsection{Chemical network}

% Description of the model and grain distribution parameters
In order to calculate the diffusivities, we use the equilibrium chemical network of \citet{ZhaoEtAl2018a}, which includes 21 major neutral species observed in dense molecular clouds, 31 corresponding ion species, electrons, and neutral and singly charged grain species. The network also contains over 500 reactions including gas-phase reactions, recombination of charged species on grains as well as freeze-out onto and thermal desorption of molecules off grains, and all possible charge-transfer reactions involving grains. This is important for obtaining correct ion abundances in the regimes where grains are the dominant charge carriers \citep{ZhaoEtAl2018a}.

We use a number of different grain size distributions in this paper, which all employ the standard -3.5 power law as in the Mathis-Rumpl-Nordsieck \citep[MRN;][]{MathisEtAl1977} distribution, but with a varying minimum and maximum grain radius, $\amin$ and $\amax$ respectively. The total grain mass is fixed at $q_{\rm g} = 1$ per cent of the gas mass, and the grain density is set as $\rho_{\rm g} = 3.0$ g cm$^{-3}$ \citep{KunzMouschovias2009}, i.e. the average density of silicates. The size range is divided logarithmically into 20 size bins between $\amin$ and $\amax$. The radial size distribution function is therefore given by
\begin{equation}
\frac{{\rm d}n(a)}{{\rm d}a} = C a^{-3.5},
\end{equation}
where the normalization factor $C$ can be determined as
\begin{equation}
C = \frac{3q_{\rm g} m_{\rm H}}{4 \pi \rho_{\rm g} \left( 
\amax^{0.5} - \amin^{0.5} \right)} n({\rm H}_2).
\end{equation}
This results in a grain surface area density of
\begin{equation}
S_{\rm g} =   \frac{6 q_{\rm g} m_{\rm H}}{\rho_{\rm g} \sqrt{\amin \amax}} n({\rm H}_2) \mbox{ cm}^{-1}. \label{eqn:surfgrains}
\end{equation}
Two of our grain distributions also include an additional single size population of PAHs, with size $a_{\rm PAH} = 0.5$ nm. The PAH abundance in T-Tauri stars estimated by \citet{GeersEtAl2006} is about $x_{\rm PAH} \approx  10^{-8} - 10^{-7}$. \citet{Bai2011b} found that above the critical PAH abundance of $x_{\rm PAH} = 10^{-9}$, the presence of tiny grains becomes significant in modifying the mass accretion, hence we adopt $x_{\rm PAH} = 10^{-8}$.

%-------------------------------------------------
% Non-ideal MHD
%-------------------------------------------------
\subsection{Non-ideal MHD}

The two non-ideal MHD coefficients relevant to this study, Ohmic and ambipolar, can be expressed in terms of the components of the conductivity tensor $\sigma$ \citep[e.g.][]{Wardle2007}:
\begin{equation}
\etaO = \frac{c^2}{4 \pi \sigma_{\rm O}}, \label{eqn:nonideal1}
\end{equation}
\begin{equation}
\etaA = \frac{c^2}{4 \pi \sigma_{\perp}} \frac{\sigma_{\rm P}}{\sigma_{\perp}} - \etaO; \label{eqn:etaAnew}
\end{equation}
where $\sigma_{\perp} = \sqrt{\sigma_{\rm H}^2 + \sigma_{\rm P}^2}$, and the Ohmic $\sigma_{\rm O}$, Hall $\sigma_{\rm H}$, and Pedersen $\sigma_{\rm P}$ conductivities are related to the Hall parameter $\beta_{i, {\rm H}_2}$:
\begin{eqnarray}
\sigma_{\rm O} = \frac{e c n\left({\rm H}_2\right)}{B} \sum_i Z_i x_i \beta_{i, {\rm H}_2}, \\
\sigma_{\rm H} = \frac{e c n\left({\rm H}_2\right)}{B} \sum_i \frac{Z_i x_i \beta_{i, {\rm H}_2}}{1 + \beta_{i, {\rm H}_2}}, \\
\sigma_{\rm P} = \frac{e c n\left({\rm H}_2\right)}{B} \sum_i \frac{Z_i x_i}{1 + \beta_{i, {\rm H}_2}}; \label{eqn:nonideal2}
\end{eqnarray}
\citep{WardleNg1999, ZhaoEtAl2016}, where $x_i$ is the abundance of charged species $i$ with respect to H$_2$ molecules and $Z_i e$ its charge, $c$ is the speed of light, and $n\left({\rm H}_2\right)$ is the number density of H$_2$. The Hall parameter determines the relative importance of the Lorentz and drag forces in determining the direction of drift for each charged species $i$. It is defined as
\begin{equation}
\beta_{i, {\rm H}_2} = \left(\frac{Z_i e B}{m_i c}\right) \frac{1}{\mu m_{\rm H} n({\rm H}_2) \gamma_i},
\end{equation}
where $m_{\rm H}$ is the mass of a hydrogen atom, and $\mu = 2.36$ is the mean molecular weight per hydrogen atom (assuming a mass fraction of 71 per cent hydrogen, 27 per cent helium, and 2 per cent metals).

For each of the different grain distribution models used in this paper, we produce a chemical look-up table providing the abundances $x_i$ for charged ion and grain species on a 3D grid in $\left[\rho, T,\zeta\right]$ space. These tables are referenced from within the simulation and are used to update the non-ideal MHD coefficients via equations (\ref{eqn:nonideal1}) - (\ref{eqn:nonideal2}) at each point in the computational domain.

%=================================================
% Grain distributions and ionization structure
%=================================================
\section{Grain distributions and ionization structure} \label{sec:diffStruct}

\begin{table}
\caption[]{Summary of simulation parameters. The symbols $\amin$ and $\amax$ represent the minimum and maximum grain radius and $\beta$ is the plasma-$\beta$.}
\begin{center}
\begin{tabular}{l c c c c}
\hline
\hline
Label & $\beta/10^3$ & $\amin$ & $\amax$ & PAHs \\
& & ($\mu \rm m$) & ($\mu \rm m$) & \\
\hline
S18-b3-A & $0.922$ & - & - & - \\
MRN-b3-A & $0.922$ & 0.005 & 1 & - \\
trMRN-b3-A & $0.922$ & 0.1 & 1 & - \\
eMRN-b3-A & $0.922$ & 1 & 100 & - \\
MRN-PAH-b3-A & $0.922$ & 0.005 & 1 & $\checkmark$ \\
trMRN-PAH-b3-A & $0.922$ & 0.1 & 1 & $\checkmark$ \\
\\
S18-b3-AO & $0.922$ & - & - & - \\
MRN-b3-AO & $0.922$ & 0.005 & 1 & - \\
trMRN-b3-AO & $0.922$ & 0.1 & 1 & - \\
eMRN-b3-AO & $0.922$ & 1 & 100 & - \\
MRN-PAH-b3-AO & $0.922$ & 0.005 & 1 & $\checkmark$ \\
trMRN-PAH-b3-AO & $0.922$ & 0.1 & 1 & $\checkmark$ \\
\\
MRN-PAH-b2-A & $0.0922$ & 0.005 & 1 & $\checkmark$ \\
MRN-PAH-b4-A & $9.22$ & 0.005 & 1 & $\checkmark$ \\
\hline
\end{tabular}
\end{center}
\label{tab:sims}
\end{table}

\begin{figure}
	\centering
	\includegraphics[width=85.5mm]{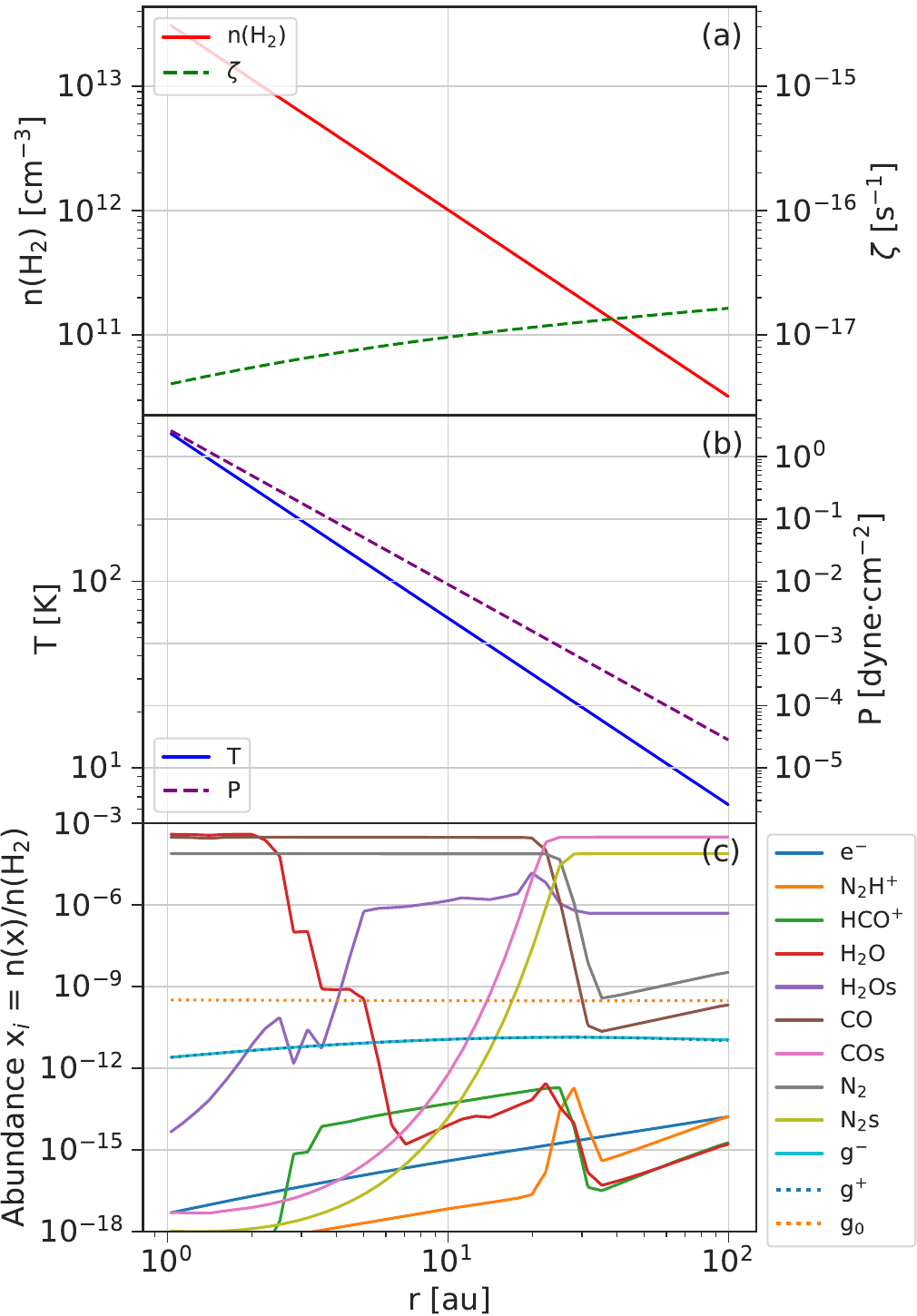}
	\caption[]{Initial mid-plane radial profiles of key disc properties. Panel (a) gives the mid-plane molecular hydrogen number density $n(H_2)$, and the total ionization rate $\zeta$, and panel (b) shows the mid-plane temperature $T$ and pressure $P$. Finally, panel (c) shows a selection of key chemical abundance ratios with respect to molecular hydrogen for the MRN grain distribution. Species in panel (c) with a trailing `s' denote molecules frozen out on the surface of grains, while $g^{\pm}$ and $g_0$ denote singly-charged and neutral grain species, respectively.}
	\label{fig:structInit}
\end{figure}

\begin{figure*}
	\centering
	\includegraphics[width=178mm]{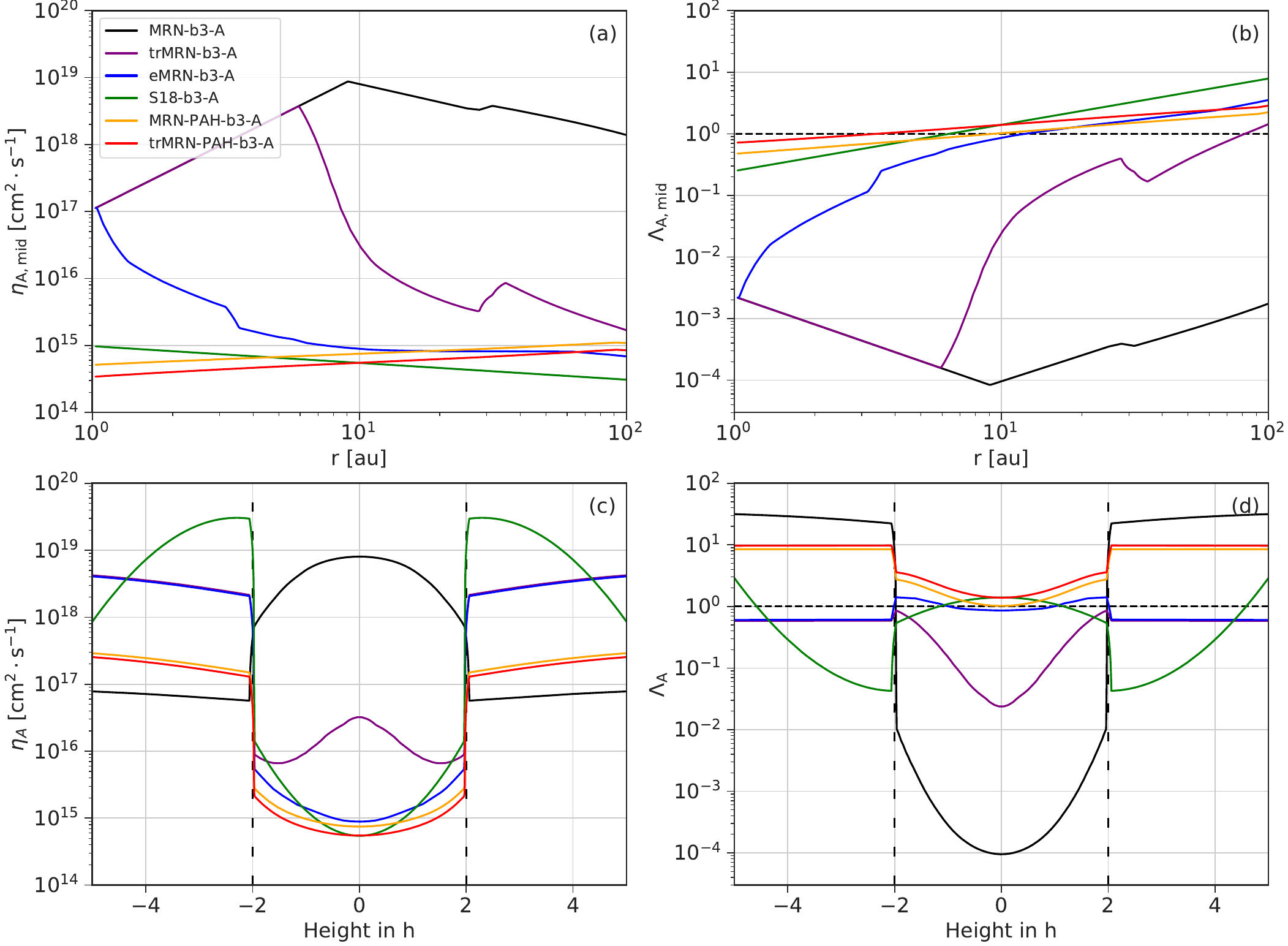}
	\caption[]{Initial profiles for the ambipolar diffusion $\etaA$ and neutral-field coupling $\ElsasserA$ for the five different grain size distributions and the reference simulation of \citet{SurianoEtAl2018}. Panels (a) and (b) show the radial profiles of $\etaA$ and $\ElsasserA$ along the disc mid-plane at $t = 0$, while panels (c) and (d) show these two parameters as a function of disc scaleheight $h$ within the disc at $r = 10$ au. The horizontal dashed lines in panels (b) and (d) show the threshold in $\ElsasserA$ between weak field-matter coupling ($\ElsasserA < 1$) and strong field-matter coupling ($\ElsasserA > 1$). The vertical dashed lines in panels (c) and (d) mark the edges of the disc, as prescribed in Section \ref{sec:method}. Note: the radial $\etaA$ and $\ElsasserA$ profiles in panels (a) and (b) are modified for $r < 9$ au due to the minimum density condition described in Section \ref{sec:bound}. Also, in panels (c) and (d) profiles for trMRN-b3-A and eMRN-b3-A overlap above the disc edge ($z = \pm 2 h$).}
	\label{fig:diffInit}
\end{figure*}

In this section, we look at how the initial disc ionization structure changes with grain distribution. We run five simulations with different grain distributions, as well as a reference simulation identical to the reference model ad-els0.25 of \citet{SurianoEtAl2018}. The parameters for all simulations are listed in Table \ref{tab:sims}. All simulations are initialized with a density at $r_0$ on the disc mid-plane of $\rho_0 = 1.265\times 10^{-10}$ g cm$^{-3}$, and poloidal magnetic field strength of $B_{\rm p,0} = 1.384 \times 10^{-1}$ G, where $r_0 = 1$ au. This fixes $\beta \sim 10^{3}$ along the entire mid-plane for all simulations, while also setting $\Lambda_0 = 0.25$ at the inner boundary for the reference simulation. We run all simulations for $5000 t_0$, where $t_0 = 1$ yr is the orbital period at $r_0$.

% Describing the different chemical models, their motivation, and where they fit in the context of PPD's.
Each of the five chemical models vary in terms of grain size distribution: a) an extended MRN distribution (denoted MRN) with $\amin = 0.005 \mu$m and $\amax = 1 \mu$m, to include larger grains expected to be present within protoplanetary discs, b) a truncated form of distribution (a) (denoted trMRN) with $\amin = 0.1 \mu$m and $\amax = 1 \mu$m, c) an evolved MRN (eMRN) distribution with $\amin = 1 \mu$m and $\amax = 100\mu$m, d) an extended MRN distribution plus a population of PAHs (MRN-PAH), and e) a truncated MRN distribution including PAHs (trMRN-PAH). We choose a wide spread of possible grain populations within the disc in order to better grasp the conditions conducive to ring and gap formation. 

%-------------------------------------------------
% MRN, trMRN and eMRN models
%-------------------------------------------------
\subsection{MRN, trMRN and eMRN models}

% The size sequence models (justification, expectations, then results)
The first three models (MRN, trMRN, eMRN) contain a general progression from smaller to larger grains. 
% Justification
The MRN distribution is based on that observed in the ISM and may represent the very earliest stages of disc formation before grain processing occurs, although thick icy mantles are known to be present before the formation of a protostar and its protoplanetary disc \citep{CaselliEtAl2022}. 
During the process of dense cloud core contraction, \citet{SilsbeeEtAl2020} showed that very small grains can be eliminated quickly, motivating our inclusion of the trMRN distribution as a possible grain distribution for early phase PPDs (Class 0). The eMRN becomes appropriate for later stages where grains have grown to sub-millimeter sizes \citep[Class II, e.g.][]{DAlessioEtAl2001, vanBoekelEtAl2003, Liu2019}.
% Expectations on diffusivities
Equation (\ref{eqn:surfgrains}) demonstrates that if the grain population is shifted to larger sizes, the grain surface area density $S_{\rm g}$ is reduced. As a result, grains are less likely to soak up free electrons, increasing the ionization fraction and lowering the amount of magnetic diffusion $\etaA$. Hence we expect that the progression \mbox{MRN $\rightarrow$ trMRN $\rightarrow$ eMRN} displays a reduction in $\etaA$, and corresponding increase in the neutral-field coupling $\ElsasserA$.

% Abundances
Panels (a) and (b) of Figure \ref{fig:structInit} show the initial conditions for all models along the disc mid-plane, with panel (c) giving fractional abundances for the MRN grain distribution based on these values. Moving outwards in radius, the temperature and density within the disc decrease, resulting in a change in chemical composition. When the temperature drops below 150 K at $r\sim 3.5$ au, H$_2$O freezes out onto the surface of dust grains \citep[e.g.][]{FuruyaAikawa2014}, with a corresponding increase in $\hcop$ abundance, as $\hcop$ is destroyed by gas-phase water \citep{PhillipsEtAl1992, BerginEtAl1998, IleeEtAl2011, LeemkerEtAl2021}. 
Similarly, outwards of $r \sim 20$ au, CO freezes out, with a corresponding jump in $\ntwohp$. N$_2$ and H$_3^+$ are the main reactants forming $\ntwohp$, but H$_3^+$ preferentially transfers a proton to CO when CO is in the gas phase. Hence outside the CO snow line, $\ntwohp$ abundance grows significantly \citep{QiEtAl2013}. A little further out at $r \sim 25$ au, N$_2$ freeze-out occurs, reducing the amount of N$_2$ available for production of $\ntwohp$. Hence, the CO and N$_2$ snow lines delimit a band of enhanced $\ntwohp$ \citep{QiEtAl2019}.

% Radial results
The chemical abundances and grain populations contribute to the magnetic diffusivities via equations (\ref{eqn:nonideal1}) - (\ref{eqn:nonideal2}). Figure \ref{fig:diffInit} shows the initial radial profiles of $\etaA$ and $\ElsasserA$ at the disc mid-plane ($\eta_{\rm A, mid}$ and $\Lambda_{\rm A, mid}$, panels a and b respectively), and between $\pm$ 5 disc scaleheights $h$ at a radius of 10 au (panels c and d). In panel (a), we see a clear drop in the mid-plane ambipolar diffusion $\eta_{\rm A, mid}$ between simulations MRN-b3-A, trMRN-b3-A and eMRN-b3-A as expected for larger grains, while $\Lambda_{\rm A, mid}$ (panel b) increases, given its inverse relationship to $\eta_{\rm A, mid}$ (see equation \ref{eqn:lambdaa}). This trend also pervades the entire height of the disc initially ($-2h \le z \le 2h$). In panels (a) and (b), the time-step limiter discussed in Section \ref{sec:bound} enforces a maximum $\etaA$ (minimum $\ElsasserA$) inwards of $r < 9$ au, allowing the simulation to proceed at a practical pace.

The H$_2$O and CO snow lines can be seen in the $\etaA$ profiles at $r \sim 3.5$ and 30 au respectively in panel (a) of Figure \ref{fig:diffInit}. The jumps in $\etaA$ for models MRN, trMRN and eMRN are due to the increased abundance in $\hcop$ between the two radii, as $\hcop$ is the dominant molecular ion in PPDs \citep[e.g.][]{TeagueEtAl2015} and hence an increased $\hcop$ abundance leads to a larger ionisation fraction and lower $\etaA$\footnote{Although ionized grains (g$^\pm$) have larger abundances than $\hcop$, they are not causing the variations in $\etaA$.}. Interestingly, there is no diffusivity bump at $\sim 30$ au for the eMRN model. This can be attributed to the order of magnitude lower grain surface area density $S_g$ compared to the other models, effectively suppressing freeze out.

%-------------------------------------------------
% PAH models
%-------------------------------------------------
\subsection{PAH models} \label{sec:PAHmodels}

% The PAH models (justification of our use of PAHs, expectations for the diffusivities, then results in plot)
For the fourth and fifth models, we add a population of PAHs to both the MRN and trMRN distributions (MRN-PAH, trMRN-PAH). 
% Justification
In these two cases, we assume that fragmentation of grains within the underlying distributions allow for the build up of PAH populations within the disc. While there is observational support for grain growth to micron size or larger in PPDs, PAH emission has been detected in the majority of Herbig Ae/Be stars \citep{AckevandenAncker2004}, as well as a small fraction of T-Tauri stars \citep{GeersEtAl2006, OliveiraEtAl2010}.

% Results
The addition of PAHs dramatically reduces $\etaA$, both along the mid-plane and with height in the disc when comparing to the two original distributions in Figure \ref{fig:diffInit}. 
% Expectations
Originally it was thought that PAHs would have the opposite effect, as they provide an immense surface area for the recombination of free electrons. However, tiny grains have been shown to greatly reduce $\etaA$ for plasmas of sufficiently weak ionization, as the main charge carrier switches from ions and electrons to grains \citep{Bai2011b}. Charged PAHs are much lighter than charged large grains and, thus, better coupled to the magnetic field. They act more like metal or molecular ions in their contribution to magnetic coupling.

This also minimises the effect of snow lines on the radial $\etaA$ profile, as ionized grains are far more abundant than ions and electrons.

% Comparison to SurianoEtAl2018
Comparing each of the grain distributions listed above with the reference simulation of \citet{SurianoEtAl2018}, we find that it at least initially compares best with the evolved MRN distribution eMRN, and the two distributions including PAHs. Whether this results in a similar disc and wind structure at later times is the subject of the next section.

%-------------------------------------------------
% Disc evolution
%-------------------------------------------------
\section{Disc/wind evolution} \label{sec:discEvol}

\begin{figure*}
	\centering
	\includegraphics[width=178mm]{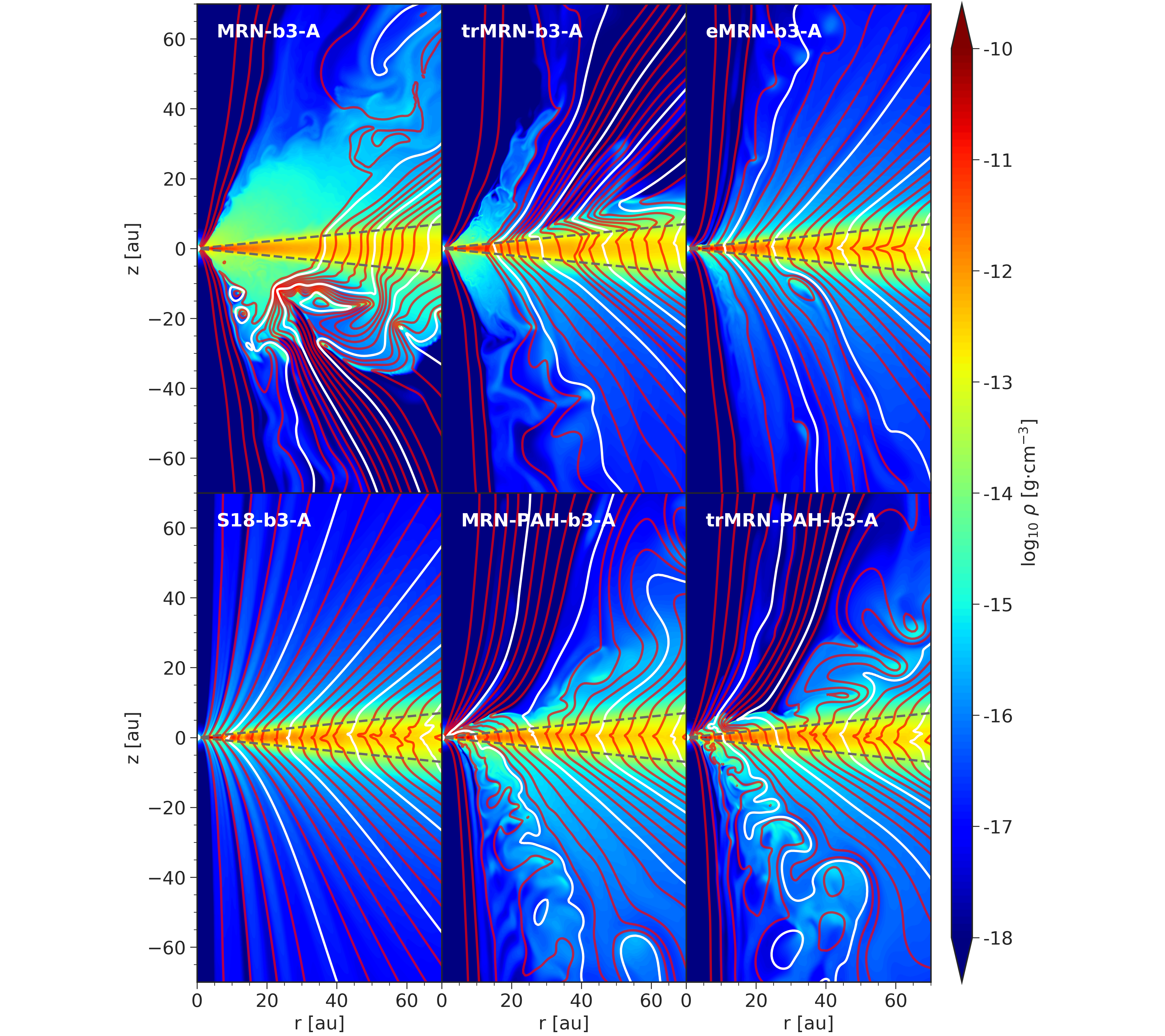}
	\caption[]{Snapshots at $t/t_0 = 2500$ from the five simulations with different grain size distributions, and the reference simulation S18-b3-A (see Table \ref{tab:sims}). Displayed are the density and the poloidal magnetic field lines (red, with periodic white contours for reference). The magnetic flux contours which show the poloidal field morphology are at the same levels for all plots, to compare the transport of field lines between simulations. We also demarcate the extent of the initial disc with dotted grey lines at $\theta = \pi/2 \pm \theta_0$. The simulation names are displayed in the top left corner of each plot.}
	\label{fig:discEvol}
\end{figure*}

\begin{figure*}
	\centering
	\includegraphics[width=178mm]{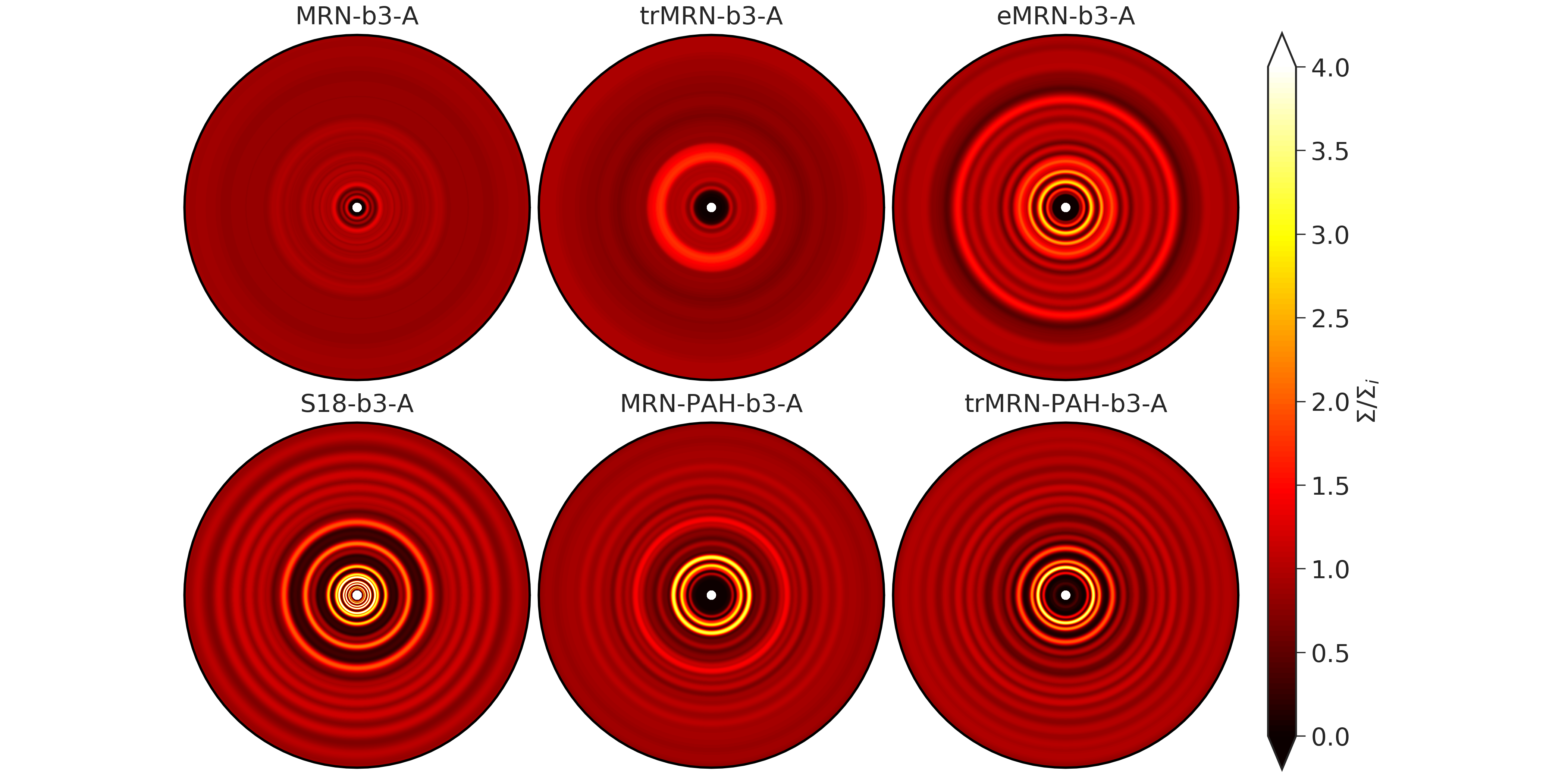}
	\caption[]{Face on surface density profiles normalized to their initial radial distribution (out to a radius of 35 au) for the five simulations with different grain size distributions and reference simulation S18-b3-A at $t/t_0 = 2500$.}
	\label{fig:discSD}
\end{figure*}

% Introduction
We now look at how the different grain distributions affect the disc and wind structure, including the radial distribution of rings and gaps. Figure \ref{fig:discEvol} displays the density distributions and magnetic field morphologies for the five simulations with different grain populations and the comparison model S18 at $t/t_0 = 2500$, while Figure \ref{fig:discSD} shows the normalized surface density profiles for $r < 35$ au at this time. We organise the panels in order of increasing initial Elsasser number at the mid-plane inner boundary $\Lambda_{\rm A,0}$, as seen in Figure \ref{fig:diffInit}, panel (b).

% Field/wind morphology - MRN model
We begin with model MRN-b3-A. 
% Magnetic field evacuation and dense atmosphere
In this model the poloidal magnetic field is asymmetric and quite sparse for $r < 35$ au, while the atmosphere surrounding the disc is dense (Figure \ref{fig:discEvol}). Due to the large grain surface area density from small grains, the initial coupling within the disc is $\ElsasserA \approx 10^{-3}-10^{-4}$ (see Figure \ref{fig:diffInit}). When the Elsasser number is this low, the ions collide with the bulk fluid so infrequently that the magnetic field is easily decoupled from the disc and transported radially outwards, leaving the inner disc with relatively little magnetic flux. 
% Dense atmosphere
This allows the disc material to expand vertically due to the absence of sufficient magnetic compression.
% Rings and gaps similarly
Likewise, significant ring growth is not seen for the MRN model (Figure \ref{fig:discSD}), as high levels of diffusion significantly reduce the ability of the magnetic field to form any notable structures within the disc.
% Accretion streams
In the southern hemisphere of the MRN model (Figure \ref{fig:discEvol}) we observe the so-called `avalanche-accretion streams' \citep{MatsumotoEtAl1996, KudohEtAl1998, SurianoEtAl2017, SurianoEtAl2018}: features characterised by highly-pinched poloidal field lines which funnel material and reconnected magnetic flux along or above the disc surface radially inwards to the inner disc, and render the disc-wind system more chaotic.

% The trMRN and eMRN simulations
The next two models (trMRN and eMRN) show increasingly stable, axisymmetric flows with pinched poloidal type magnetic field configurations. The trMRN model exhibits an asymmetric configuration, similar to that shown in model MRN-b3-A. In the southern hemisphere, the classic swept-back magnetic field configuration develops quickly, and is maintained for the duration of the simulation. In the northern hemisphere however, avalanche streams funnel material up from the outer disc ($r \sim 80$ au) and radially inwards above the disc surface towards the inner 10 au. The wind launched from the inner disc competes with these avalanche streams, creating a cyclic pattern of wind disruption and liberation. In contrast, a very stable and nearly axisymmetric wind is set up in the eMRN model within the first 2500 yrs, coinciding with the disappearance of avalanche streams. This is also the case for the reference model S18, with the formation of a very steady, uninterrupted flow early on.
% Surface density and rings
Rings in the trMRN and eMRN models have greater contrast than the MRN model (Figure \ref{fig:discSD}) due to the overall increase in neutral-field coupling, allowing increased clumping of poloidal magnetic flux and density within the disc. One main ring is seen at $r \sim 10$ au for the trMRN model, while the eMRN model has an even greater ring/gap contrast, with periodic rings out to 20 au. In contrast, model S18-b3-A exhibits periodic rings extending from the inner boundary out to at least 35 au.

% The PAH simulations
With the addition of PAHs, the field morphology again becomes increasingly chaotic, with a smooth, swept-back wind configuration in the southern hemisphere and avalanche streams dominating the flow in the northern hemisphere (Figure \ref{fig:discEvol}). In contrast to the trMRN model however, the streams  originate from closer in within the disc at $r \sim 40-50$ au. 
% Explanation for increasingly chaotic flows
As shown in Section \ref{sec:PAHmodels}, the Elsasser number is increased by between 1--2 orders of magnitude with the addition of PAHs, which may account for the increasingly unsteady flows \citep[e.g.][]{SurianoEtAl2018}.
% Surface density and rings
The rings and gaps in both PAH models are much more periodic and radially extended than the other simulations, despite the transition to a more unstable wind state; they much more closely reflect the ring distribution of S18-b3-A than any of the others. Hence the stability of the wind does not seem to heavily influence the structure of rings in our discs.

%-------------------------------------------------
% Ring structure within the disc
%-------------------------------------------------
\newpage
\subsection{Ring structure within the disc} \label{sec:ringstruct}

\begin{figure}
	\centering
	\includegraphics[width=85.5mm]{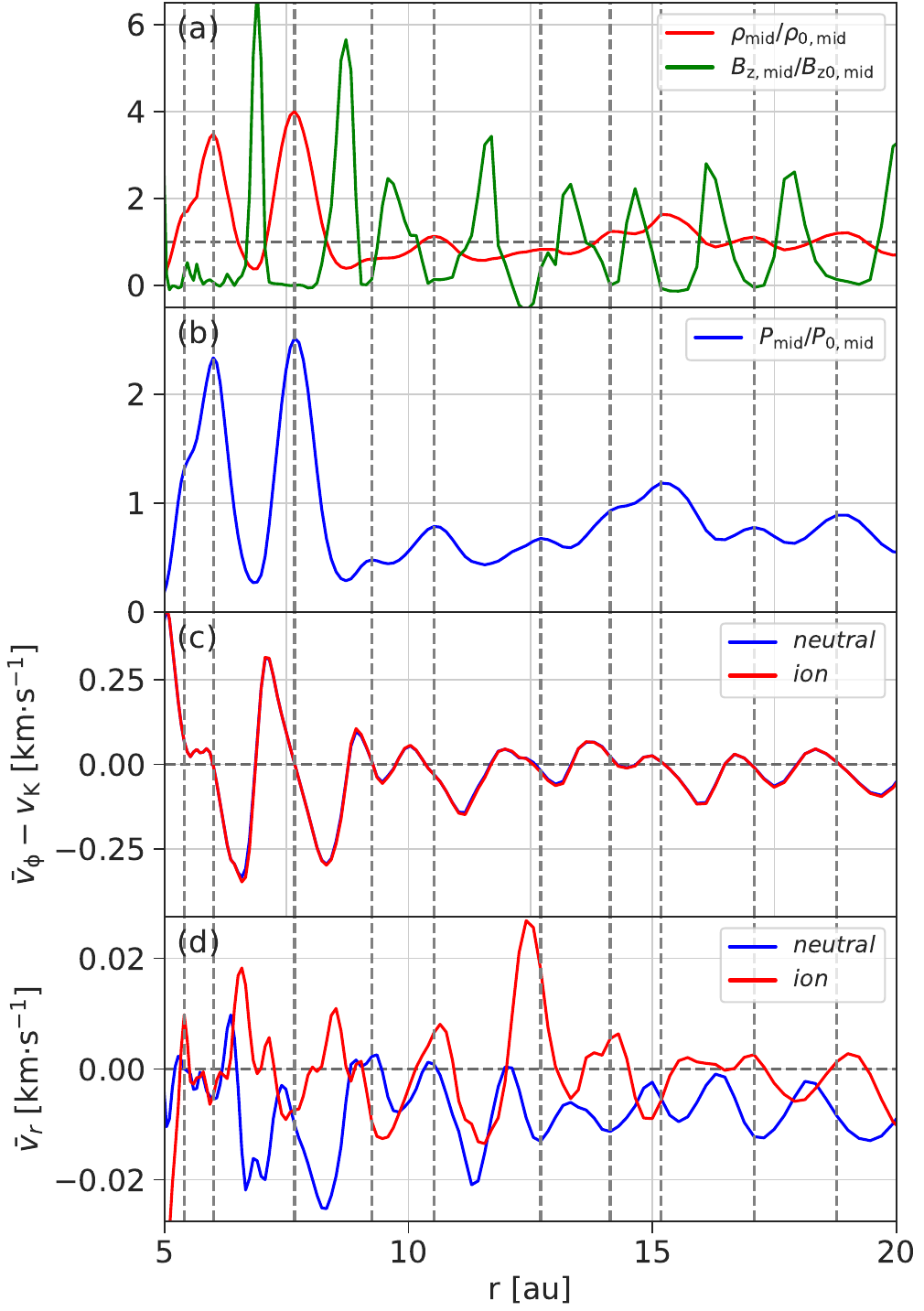}
	\caption[]{The density, magnetic field strength, pressure and velocities between 5 and 20 au for the MRN-PAH model at $t/t_0 = 2500$. The panels show (a) the mid-plane density and magnetic field strength  normalized to the their initial radial distributions, (b) the mid-plane pressure, also normalized, (c) the azimuthal velocity of neutrals and ions with respect to the Keplerian velocity, averaged over the vertical disc, and (d) the radial velocity of neutrals and ions averaged over the vertical disc. Vertical grey dashed lines denote peaks in the mid-plane density distributions (i.e. rings).}
	\label{fig:discRad}
\end{figure}

\begin{figure*}
	\centering
	\includegraphics[width=178mm]{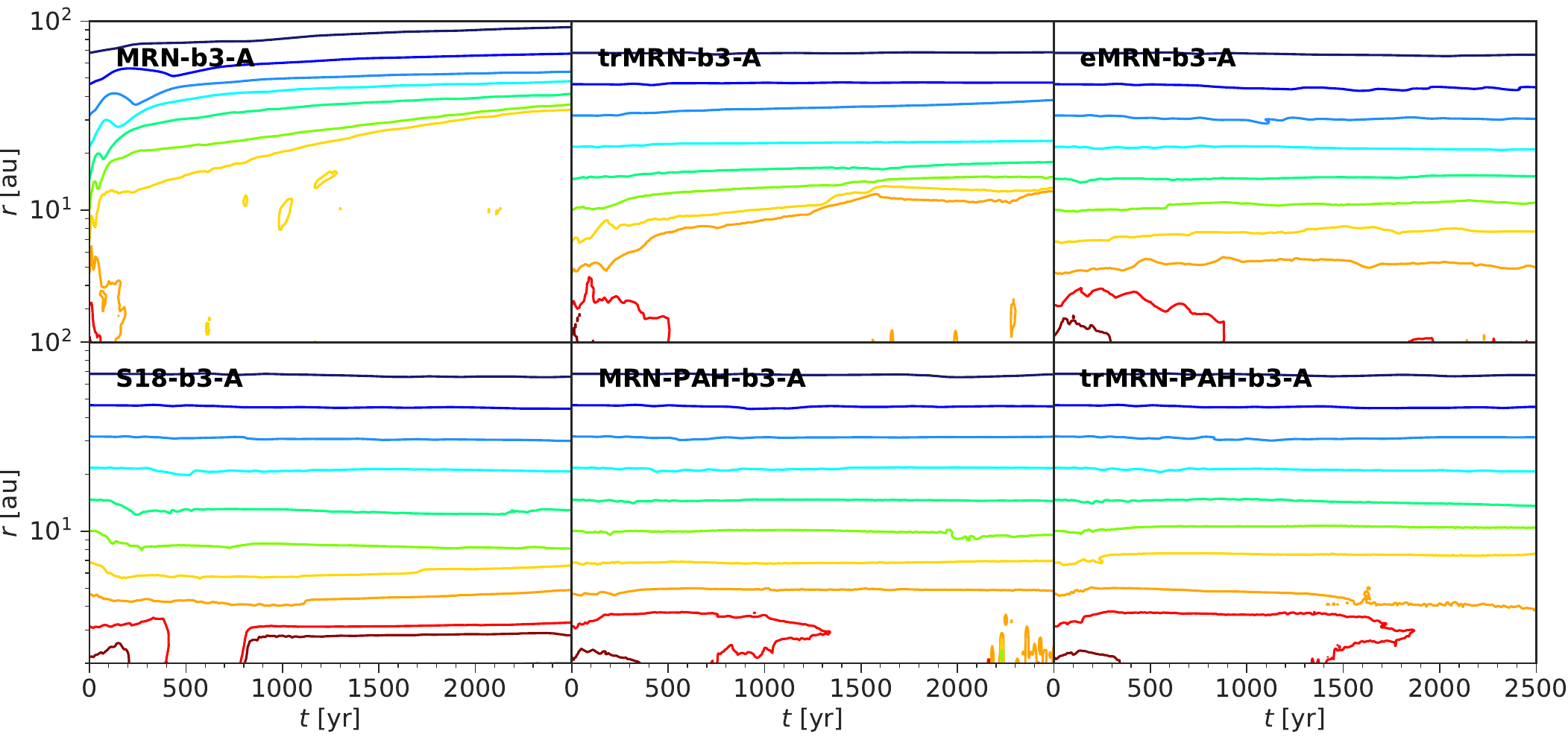}
	\caption[]{The evolution of a selection of magnetic field lines' radial intersections with the disc mid-plane as a function of time, for the five simulations with different grain size distributions and the reference simulation S18-b3-A. Initial radii are at $\left[ 2.2, \;3.2, \;4.6, \;6.8, \;10, \;15, \;22, \;32, \;46, \;68\right]$ au.}
	\label{fig:discFlux}
\end{figure*}

%Intro
We now examine the structure of rings and gaps formed within our simulations. As each of the simulations evolve, radial variations in the disc density emerge, depending on their chemical and dust composition, and become quasi-steady in their location and structure. In this paper we refer to the over-dense regions as `rings', and the evacuated regions as `gaps'. Rings and gaps in all simulations exhibit similar characteristics, and we show a selection of these from the MRN-PAH-b3-A model in Figure \ref{fig:discRad}.

% First: density and magnetic fields
The most important observation is that the formation of dense rings correlates with the collection of the magnetic flux into the low-density gaps (panel a, Figure \ref{fig:discRad}). This phenomenon is reproduced in numerous studies \citep[e.g.][]{BaiStone2014,Bai2015,SurianoEtAl2018,RiolsLesur2018,RiolsEtAl2020}, and different theories have been proposed regarding the origin of these structures. \citet{SurianoEtAl2018} propose a magnetic reconnection mechanism to produce the rings and gaps, while \citet{RiolsLesur2018,RiolsEtAl2020} suggest an instability triggered by the expulsion of wind material combined with accretion through the disc.

% Second: Pressure and azimuthal velocity structures
The second observation is that of thermal pressure bumps aligned with the dense rings (panel b). These bumps are anti-correlated with the position of the magnetic pressure bumps, resulting in a smooth total pressure within the disc. The thermally over-pressured regions modify the azimuthal velocity structure of the disc (panel c). At the inward surface of the rings, the positive thermal pressure gradient generates a region of super-Keplerian azimuthal flow, due to the positive radial pressure force, while at the outward surface, the negative pressure gradient generates sub-Keplerian azimuthal flow.

% Third: radial velocity variation
Thirdly, we see a definite variation in the radial velocity profile with radius (panel d). Maximum inward radial velocities for both neutrals and ions\footnote{We define an effective ion velocity as 
\begin{equation}
\mathbf{v}_i=\mathbf{v} - {4\pi \eta_{A}\over c}{ \mathbf{J}\times\mathbf{B}\over B^2},
\end{equation}
so the magnetic field is frozen in the effective ``ion" fluid moving with $\mathbf{v}_i$ according to the induction equation~(\ref{eqn:induction}).} correlate with gaps for $r < 12$ au, while correlating with the dense rings for $r > 12$ au. From panel (d), it is clear that there is a general tendency for the gas accretion to advect the poloidal field inward (see the blue curve in the panel). In contrast, the effective ions (and the field lines tied to them) tend to move inward at a slower speed or even expand outward because of ambipolar diffusion driven by a generally outward-directing magnetic force (see the red curve). The competition between the inward advection and outward diffusion is expected to play a key role in the magnetically-induced substructure formation by shaping the magnetic flux transport in the disk, although the exact mechanism remains unclear. 
In our simulations we observe a number of different poloidal velocity configurations with respect to the ring/gap structures. For some models we see an accretion channel in line with the mid-plane, others display a dominant accretion flow on the surface, with still others displaying a combination of the two, or a more chaotic state. This very much depends on the stability of the flow in question. 

Figure \ref{fig:discFlux} displays the evolution over time of the magnetic flux threading the disc for the six different simulations shown in Figures \ref{fig:discEvol} and \ref{fig:discSD}. In the MRN simulation, we see that all flux within 7 au accretes inwards, while the remaining flux migrates rapidly outwards. Flux initially anchored at 9 au migrates out to 35 au within 2500 years. For the trMRN simulation, this process is less extreme, with flux anchored at 4.5 au migrating out to 13 au in the same time frame. With large enough grains, and/or field-neutral coupling of $\Elsasser \gtrsim 1$ (e.g. eMRN, S18, MRN-PAH and trMRN-PAH simulations), the flux is relatively stable and we observe no outward migration.

The outward migration of flux can be attributed to the very low field-neutral coupling in both MRN and trMRN simulations (see Figure \ref{fig:diffInit}, panel b). Low coupling allows the magnetic field to flow effectively unimpeded through the gas, and combined with a positive radial magnetic field pressure the field moves rapidly outward towards a new equilibrium state.

We should note that sharp pinching of poloidal field lines near the midplane can lead to magnetic reconnection (see also \citealt{SurianoEtAl2018}), which can contribute to the outward diffusion of large-scale field lines against inward advection by the accretion flow. However, this effect is difficult to quantify.

%-------------------------------------------------
% Mass loss via accretion and winds
%-------------------------------------------------
\subsection{Mass loss via accretion and winds} \label{sec:inout}

\begin{figure}
	\centering
	\includegraphics[width=85.5mm]{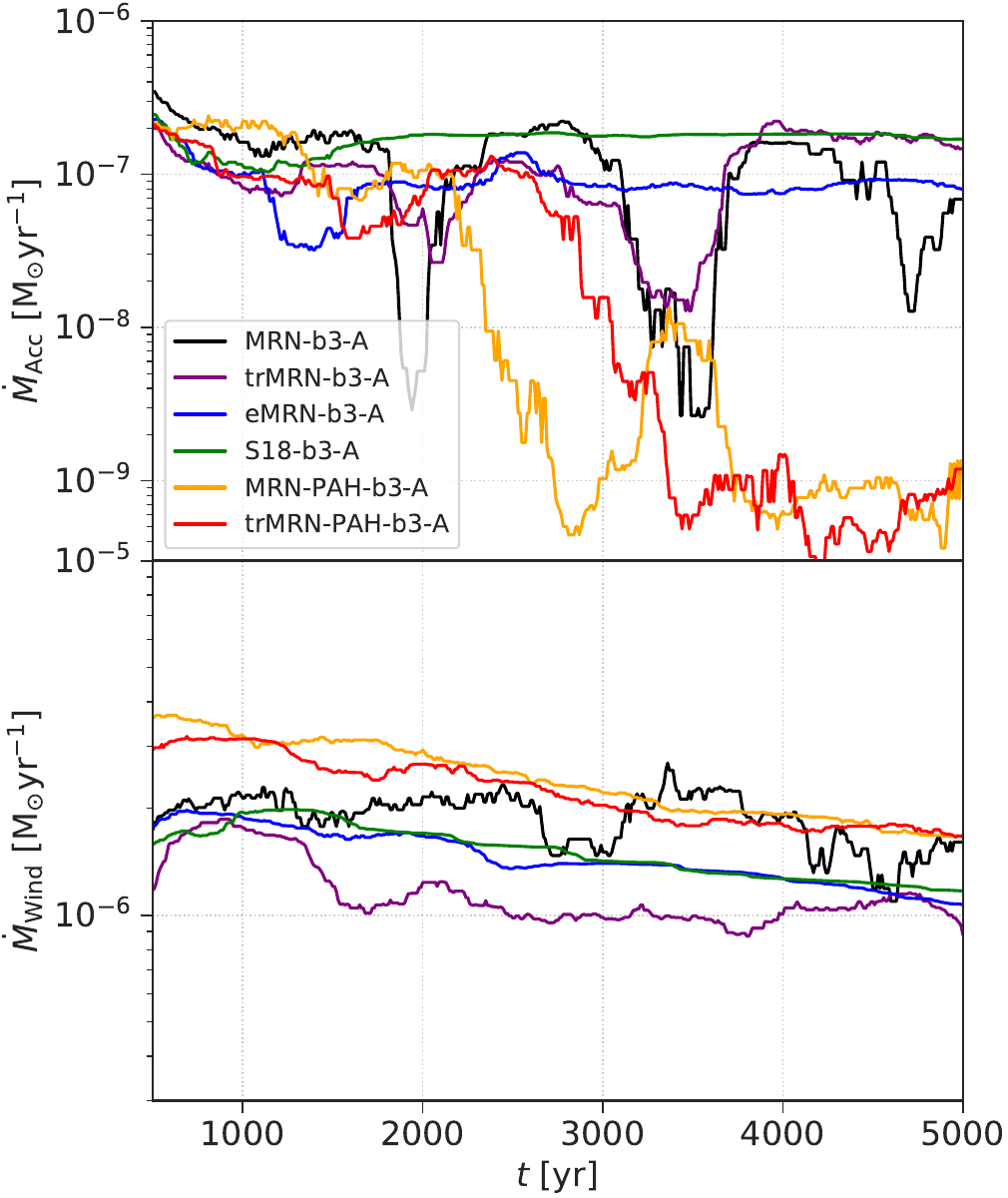}
	\caption[]{Mass accretion rates $\mdota$ and wind mass loss rates $\mdotw$ as a function of time for the five simulations with different grain size distributions, and the reference simulation S18-b3-A (see Table \ref{tab:sims}).}
	\label{fig:discAcc}
\end{figure}

% Methodology
We now investigate how the mass accretion rate $\mdota$ and wind mass loss rate $\mdotw$ change with grain size distribution. Figure \ref{fig:discAcc} plots both the mass accretion rate through the inner boundary at $r_0$, integrated between $\theta = \pi/2 \pm 2\theta_0$ at constant $\phi$:
\begin{equation}
\mdota = -2 \pi r_0^2 \int^{\pi/2 + 2\theta_0}_{\pi/2 - 2\theta_0}  \rho {\rm v}_r \sin \theta {\rm d} \theta,
\end{equation}
and the wind mass loss rate through the $r$-$\phi$ surfaces at \mbox{$\theta = \pi/2 \pm 2\theta_0$}
\begin{equation}
\mdotw = 2 \pi \left. \int^{r_{\rm out}}_{r_0} r \rho {\rm v}_{\theta} \sin \theta {\rm d}r \right]^{\theta = \pi/2 + 2\theta_0}_{\theta = \pi/2 - 2\theta_0},
\end{equation}
where $r_{\rm out} = 100$ au. The time series of $\mdota$ and $\mdotw$ are overlaid with a moving median filter of width 500 yrs 
\begin{equation}
y_{\rm MED}(t) = {\rm MEDIAN} \lbrack y(t-w), \ldots, y(t) \rbrack,
\end{equation}
where $w$ is the width of the filter, to observe the long-term trends in the flow. We should note that  $\mdotw$ captures only the sum over all disk radii of the wind mass loss, which varies strongly from one radius to another. Similarly, the disk mass accretion rate also has a strong radial variation.

Contrary to the method of \citet{SurianoEtAl2018}, which uses a spherical shell at 10 au to define $\mdota$ and $\mdotw$, our method captures the wind mass loss out to 100 au, while also only defining accretion as the material which accretes onto the `young stellar object' through the inner boundary. This explains why for the S18-b3-A model reproduced here, we report a lower $\mdota$ and higher $\mdotw$ than \citet{SurianoEtAl2018}.

% Results - Accretion
As shown in Figure \ref{fig:discAcc}, accretion rates for all models are initially \mbox{$\mdota \approx 2 \times 10^{-7}$ $\msun$ yr$^{-1}$}, but vary substantially over time. The two simulations MRN and trMRN show moderate variability after $1800$ yrs, while the eMRN and S18 simulations have roughly constant accretion rates for the full 5000 years. On the other hand, both PAH models drop down to \mbox{$\sim 10^{-9}$ $\msun$ yr$^{-1}$} after $2250 - 3500$ yrs, remaining at this rate for the remainder of the simulation. 
% Trends in variability
As shown in Section \ref{sec:diffStruct}, simulations MRN and trMRN have lower field-neutral coupling, resulting in numerous avalanche accretion streams and a dynamic mid-plane magnetic field (see Figure \ref{fig:discFlux}), which could explain the variability in $\mdota$. 
% eMRN and S18
Increasing the grain size raises the neutral-field coupling closer to unity (eMRN, S18). This level of $\Elsasser$ stabilises the magnetic field without triggering the MRI, allowing for relatively steady accretion flows and winds to operate (the blue and green lines in Fig. \ref{fig:discAcc}).
%PAH models
Adding PAHs to the MRN and trMRN grain populations further increases the neutral-field coupling above unity. This activates the MRI, resulting again in the formation of avalanche accretion streams. As shown in the last two panels of Fig. \ref{fig:discEvol}, these accretion streams can draw additional magnetic flux down towards the inner regions of the system, which can obstruct the accretion of material towards the `star', reducing $\mdota$. 

% Results - Outflow
Wind mass loss rates for all simulations remain quite steady at \mbox{$\mdotw \approx 1-4 \times 10^{-6}$ $\msun$ yr$^{-1}$}, with a slight decrease over time due to a reduction in the overall disc mass. Apart from the outlying MRN simulation\footnote{The MRN disc in Fig. \ref{fig:discEvol} is relatively puffed up compared to the other models due to the lack of magnetic pressure around the disc. This may give different results for $\mdotw$ given the integration surfaces we use.}, we also see a general increase in $\mdotw$ with grain size and the addition of PAHs. This is expected due to the increased magnetic lever arm with increased neutral-field coupling \citep[e.g.][]{AllenEtAl2003}. 

% Summary and discussion
From the results presented here, we see that accretion variability and the magnitude of $\mdotw$ are highly dependent on grain distribution. Accretion variability is low around the optimal neutral-field coupling level for steady flows (i.e. $\ElsasserA \approx 1$), with stronger or weaker coupling leading to avalanche accretion streams which disrupt the accretion flow. Weaker coupling produces a dynamic mid-plane magnetic field and overall unsteady conditions, while stronger coupling activates the MRI. $\mdotw$ correlates with neutral-field coupling, which increases with grain size and the addition of PAHs.

In summary, since the distribution of grains in the disc is a dominant factor in the determination of the neutral-field coupling, disc grain composition is a critical component in the formation of rings and the behaviour of the accretion and wind flows.

%=================================================
% Effects of ohmic diffusivity and magnetic field 
% strength on ring and gap formation
%=================================================
\section{Effects of ohmic diffusivity and magnetic field strength on ring and gap formation} \label{sec:ohmicbeta}

We now explore how the addition of Ohmic diffusion and changing the initial mid-plane plasma-$\beta$ affect the formation of rings and gaps within protoplanetary discs, and compare these results to the findings of \citet{SurianoEtAl2018}.

%-------------------------------------------------
% Ohmic diffusivity
%-------------------------------------------------
\subsection{Ohmic diffusivity} \label{sec:ohmic}

\begin{figure}
	\centering
	\includegraphics[width=85.5mm]{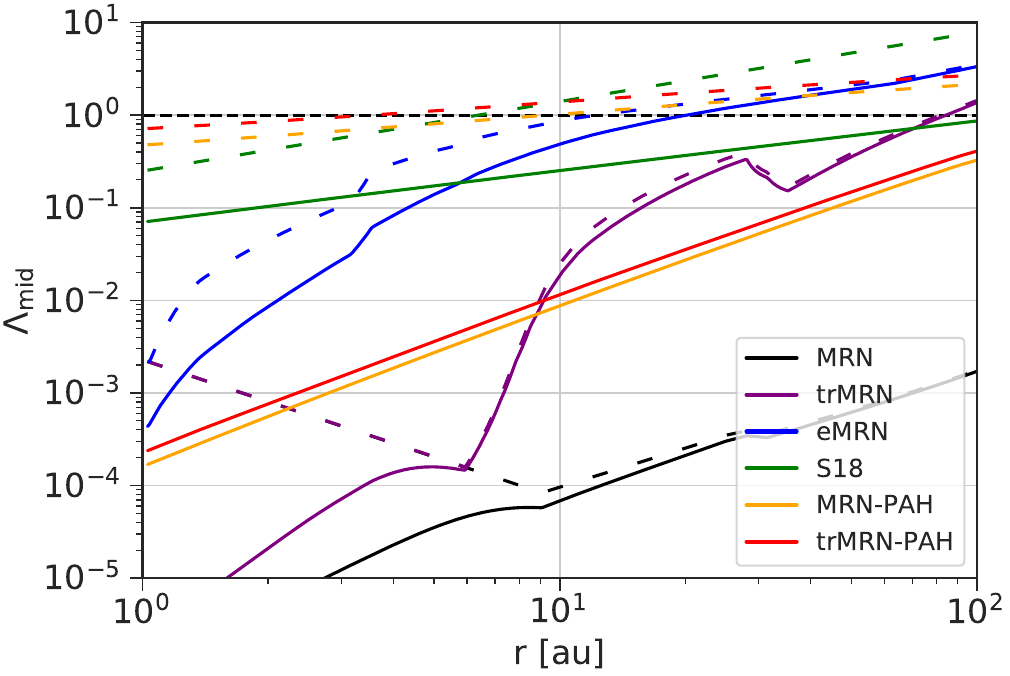}
	\caption[]{Initial mid-plane profiles for the ambipolar neutral-field coupling parameter $\ElsasserA$ (dashed lines) for the six different ambipolar-only simulations presented in Section \ref{sec:diffStruct}, and the effective neutral-field coupling parameter $\Elsasser$ (solid lines) for the six ambipolar-and-Ohmic simulations presented in Section \ref{sec:ohmic}.}
	\label{fig:discInitChem}
\end{figure}

\begin{figure*}
	\centering
	\includegraphics[width=178mm]{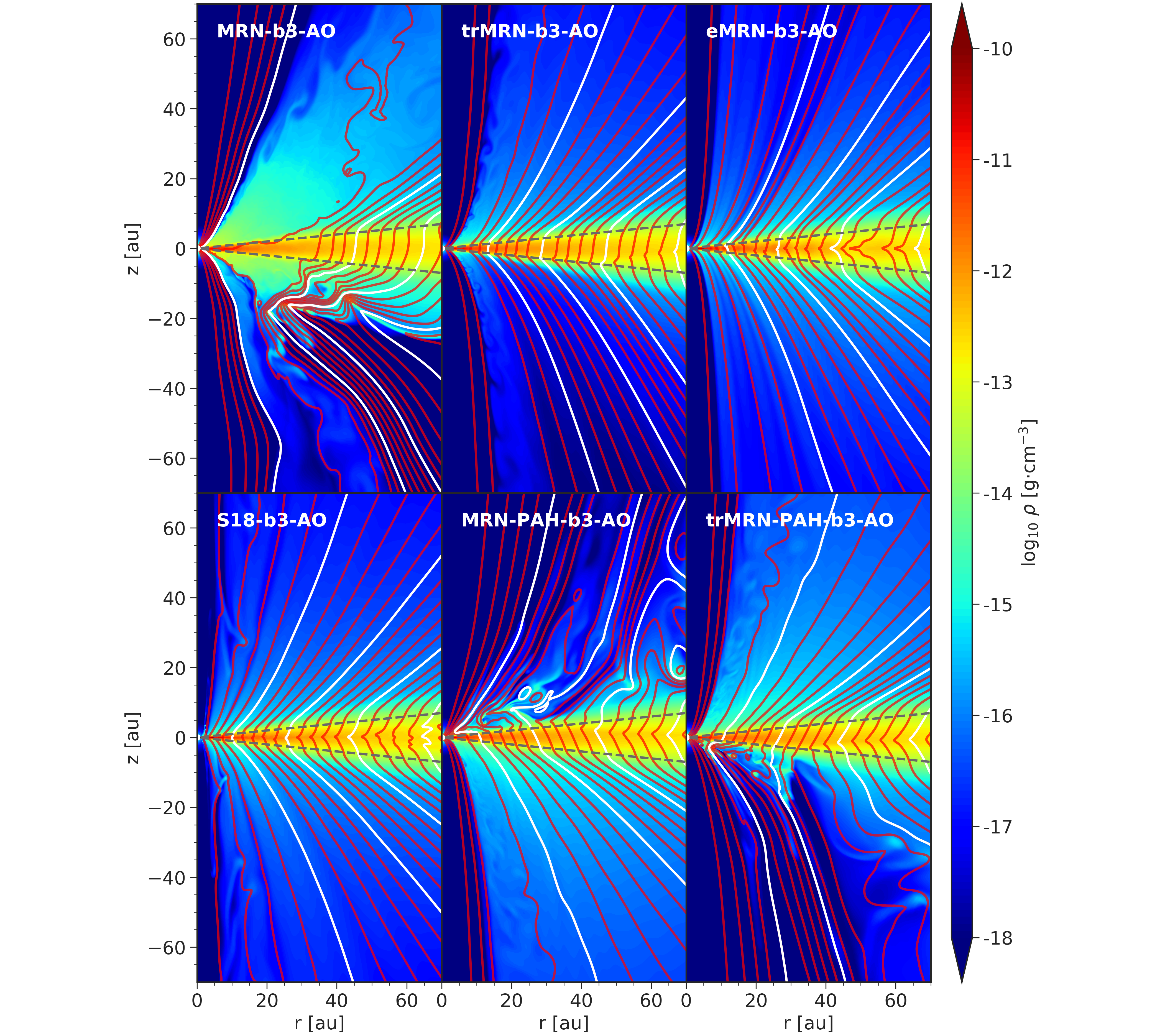}
	\caption[]{Same as Figure \ref{fig:discEvol}, but now including Ohmic diffusion as well as ambipolar.}
	\label{fig:discEvolOhm}
\end{figure*}

\begin{figure*}
	\centering
	\includegraphics[width=178mm]{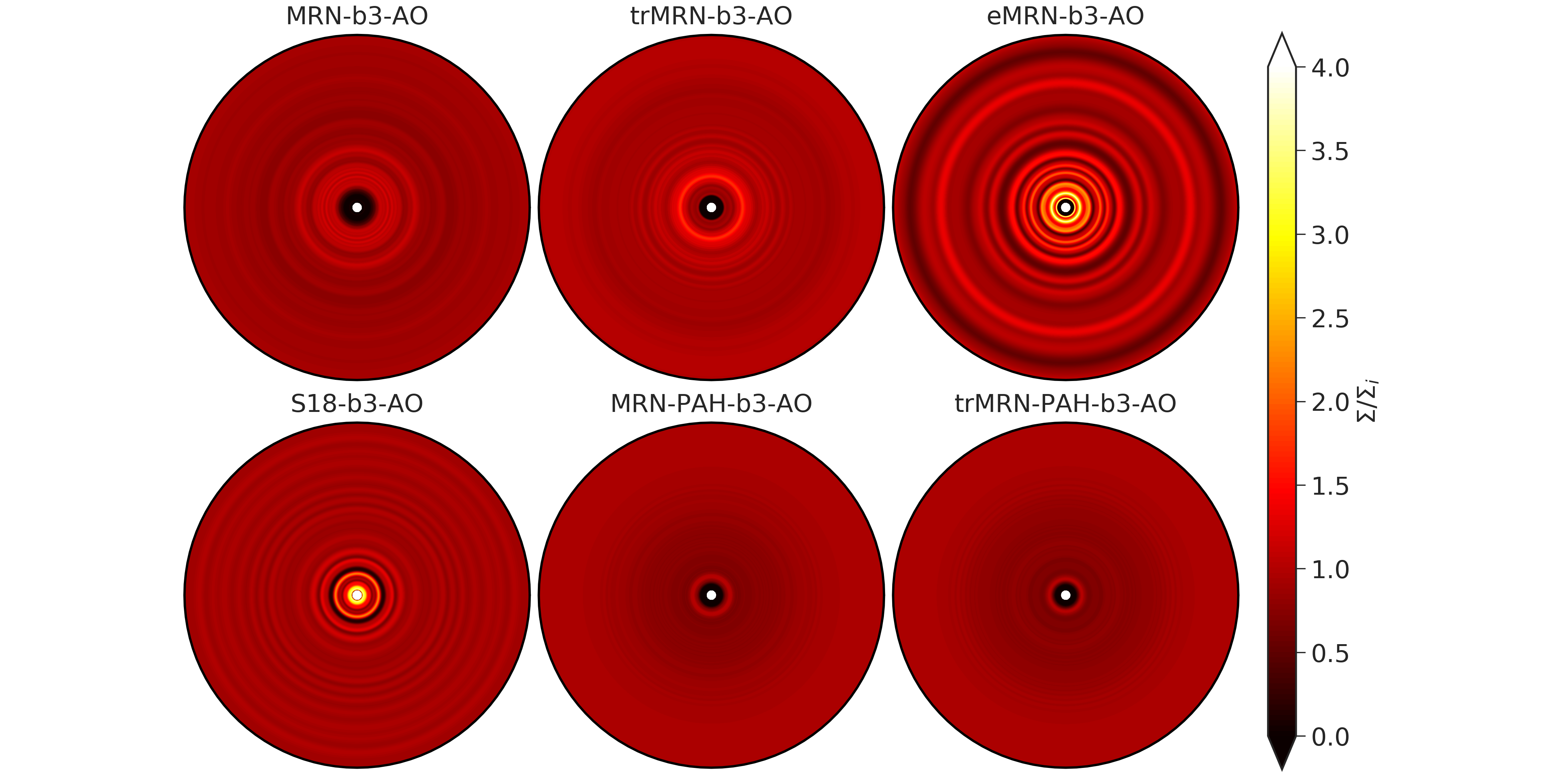}
	\caption[]{Same as Figure \ref{fig:discSD}, but now including Ohmic diffusion as well as ambipolar.}
	\label{fig:discSDOhm}
\end{figure*}

% Introduction
Ohmic diffusion is present in all protoplanetary discs, dominating at high densities near the disc mid-plane and close to the central object \citep{KoniglEtAl2010,TurnerEtAl2014}. Though not the most important effect in the outer regions of discs, Ohmic diffusion can still influence the formation of rings and gaps \citep[e.g.][]{SurianoEtAl2017}. 
% Methodology
In this section we add Ohmic diffusion to the five simulations presented in Sections \ref{sec:diffStruct} and \ref{sec:discEvol}, whilst also re-running the intermediate Ohmic model from \citet{SurianoEtAl2018} (oh2.6, now S18-b3-AO). In S18-b3-AO, the Ohmic diffusion term is set as constant everywhere in the simulation domain, and is equal to 2.6 times the initial ambipolar diffusivity $\eta_{\rm A,0}$ at $r_0$ on the disc mid-plane, translating to a value of $\etaO = 2.5 \times 10^{15}$ cm$^2$ s$^{-1}$. However, in all other Ohmic simulations we calculate the Ohmic diffusion self-consistently from the equilibrium chemical network described in Section \ref{sec:chem} and equations~(22) and (24).

% Initial coupling comparison
Figure \ref{fig:discInitChem} compares the initial mid-plane ambipolar Elsasser numbers for the ambipolar models from Section \ref{sec:diffStruct} ($\ElsasserA$, dashed), with the effective coupling for the ambipolar-and-Ohmic models shown in this section ($\Elsasser$, solid). The effective Elsasser number is defined by
\begin{equation}
\Elsasser = \frac{\ElsasserA\ElsasserO}{\ElsasserA + \ElsasserO}, \label{eqn:effectivelambda}
\end{equation} 
for ambipolar and Ohmic diffusion (see Appendix \ref{sec:devdiff} for the derivation). We see that while the addition of Ohmic diffusion has a marginal effect on $\Elsasser$ for the MRN, trMRN and eMRN grain distributions, it significantly reduces $\Elsasser$ by several orders of magnitude in the reference S18 model, and those with a PAH population (MRN-PAH and trMRN-PAH).

% Describing the global figure
Figure \ref{fig:discEvolOhm} displays the density distributions and magnetic field morphologies for the ambipolar-and-Ohmic simulations when substructures are fully developed at $t/t_0 = 2500$. Comparing to the ambipolar-only simulations in Figure \ref{fig:discEvol}, we see that with the inclusion of Ohmic diffusion, the magnetic wind morphology (with the exception of the MRN-b3-AO model) appears much more stable, with lower incidence of avalanche accretion streams and general fluctuation in the poloidal field lines.
Observing the surface density contrast for the ambipolar-Ohmic models in Figure \ref{fig:discSDOhm}, we see directly the dampening effect of Ohmic diffusion on the formation of ring structures within the disc. For the S18 model, periodic ring structures are still present with the addition of Ohmic diffusion, but the ring and gap surface density contrast is markedly reduced. In the case of the PAH simulations however, Ohmic diffusion suppresses ring formation at all but the very innermost radii. 

Comparing Figure \ref{fig:discSDOhm} to the coupling profiles in Figure \ref{fig:discInitChem}, we see a direct correlation between the values of $\Elsasser$ and the formation of stable, periodic rings in the disc. While the MRN and trMRN grain distributions are incapable of forming rings due to their very low $\Elsasser$ at most radii, for the eMRN distribution, where $\Elsasser$ remains of order unity for $r > 10$ au, rings form for both ambipolar-only and ambipolar-and-Ohmic regimes. For the S18 ambipolar-only model, $\Elsasser \gtrsim 1$ and strong periodic rings form. However when we add Ohmic diffusion, $\Elsasser$ drops an order of magnitude and while the rings remain they are no longer as prominent. Finally, for the two PAH models, $\Elsasser \gtrsim 1$ for the ambipolar-only regime, and stable, periodic rings are observed, however when Ohmic diffusion is included, $\Elsasser$ drops two orders of magnitude and the periodic rings disappear completely. Hence from these simulations we postulate that $\Elsasser \gtrsim 1$ is a general requirement for ring formation.

% Discussion
From the above results, we see the importance of self-consistent chemistry calculations in the determination of the relative diffusion magnitudes, due to their importance to the formation of rings and gaps. Both ambipolar and Ohmic diffusion levels are inextricably linked to the underlying molecular and dust-based chemistry present within the disk. While ambipolar diffusion alone may facilitate the formation of stable, periodic rings within PPDs, the additional contribution of Ohmic diffusion based on the same chemistry may suppress it. The results above also stress the importance of including reliable estimates of the dust and gas composition of the disc.

%-------------------------------------------------
% Plasma beta
%-------------------------------------------------
\subsection{Plasma beta} \label{sec:magflux}

\begin{figure}
	\centering
	\includegraphics[width=85.5mm]{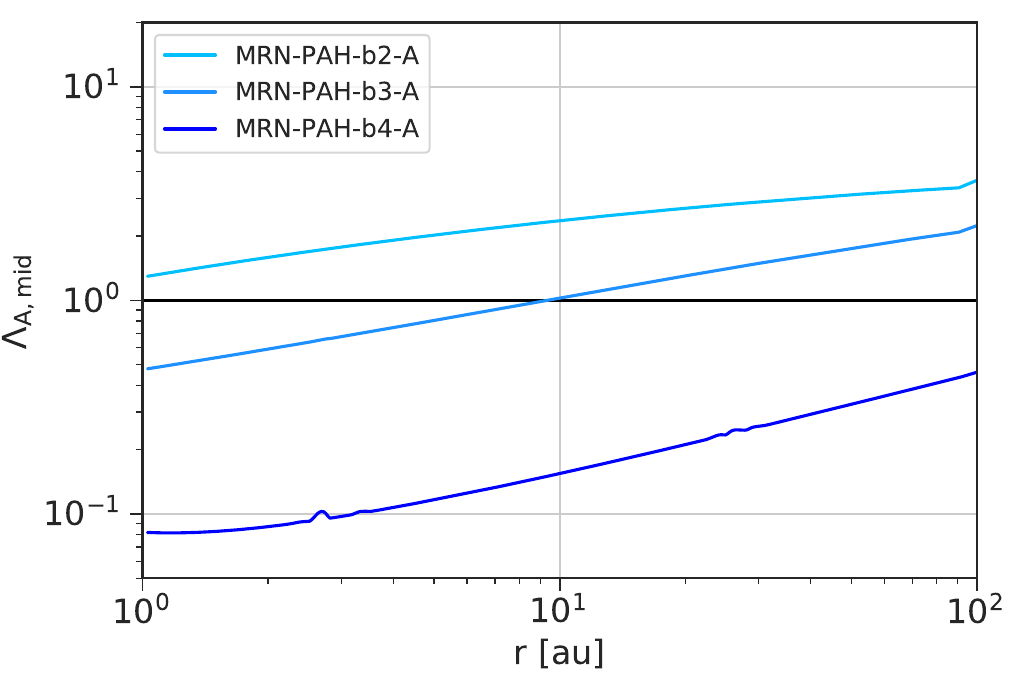}
	\caption[]{Initial mid-plane profiles for the ambipolar neutral-field coupling parameter $\ElsasserA$ for the three simulations with different plasma-$\beta$ presented in Section \ref{sec:magflux}.}
	\label{fig:discInitBeta}
\end{figure}

\begin{figure*}
	\centering
	\includegraphics[width=178mm]{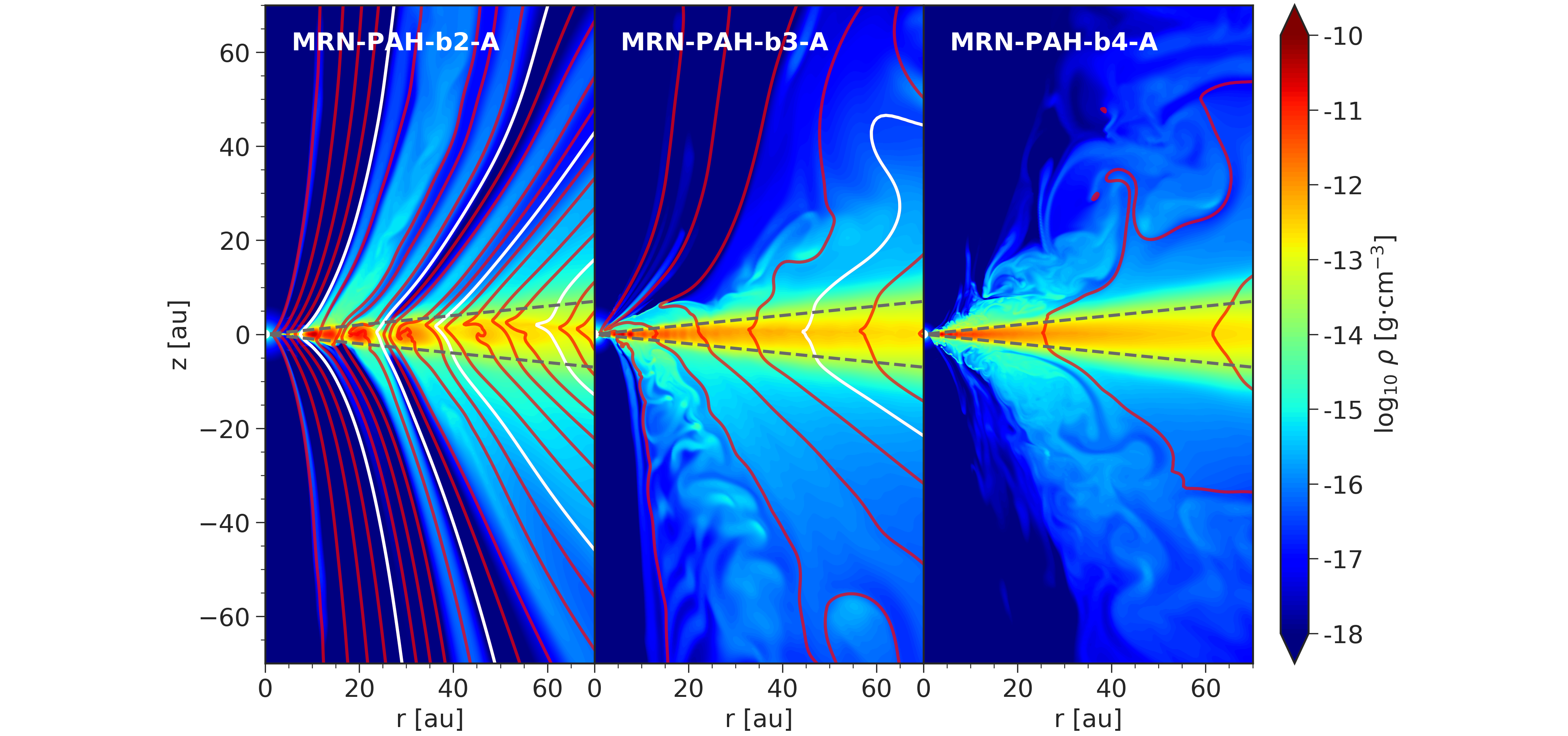}
	\caption[]{Snapshots at $t/t_0 = 2500$ from the three simulations with different plasma-$\beta$ based off the MRN-PAH-b3-A simulation (centre panel). From left to right $\beta = 10^2$, $10^3$, and $10^4$. Displayed are the density and the poloidal magnetic field lines (red, with periodic white contours for reference). The magnetic flux contours which show the poloidal field morphology are at the same levels for all plots, hence the decrease in frequency with higher $\beta$. We also demarcate the extent of the initial disc with dotted grey lines at $\theta = \pi/2 \pm \theta_0$. The simulation names are displayed in the top left corner of each plot.}
	\label{fig:discEvolBeta}
\end{figure*}

\begin{figure*}
	\centering
	\includegraphics[width=178mm]{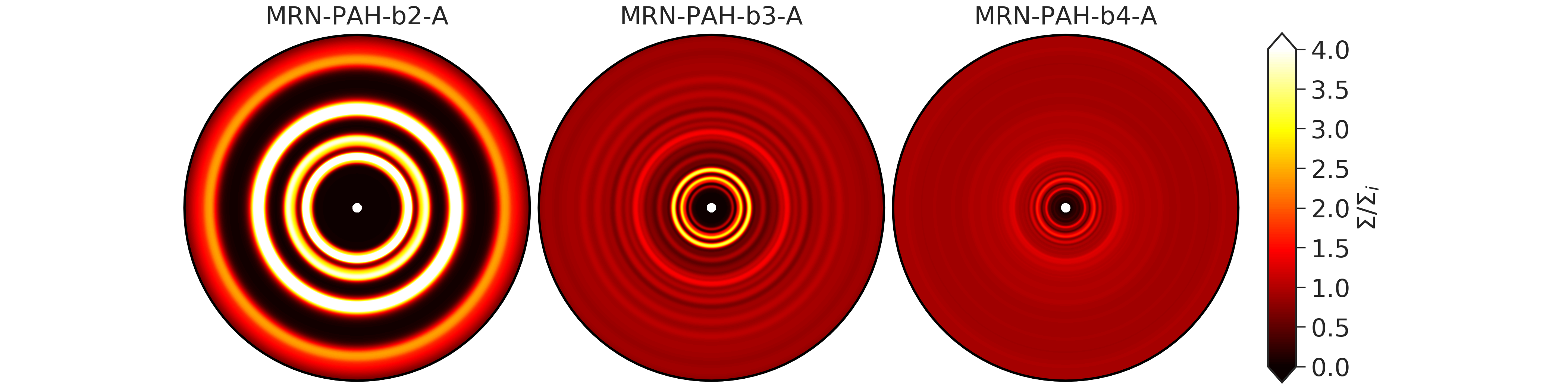}
	\caption[]{Face on surface density profiles normalized to their initial radial distribution (out to a radius of 35 au) for the three simulations with different plasma-$\beta$ presented in Figure \ref{fig:discEvolBeta} at $t/t_0 = 2500$.}
	\label{fig:discSDBeta}
\end{figure*}

% Introduction/justification
Given that the magnetic field has been shown to play such an important role in the formation of rings within non-ideal discs, we also investigate how changing the level of magnetic flux in the disc affects the ring formation mechanism. As explained in Section \ref{sec:magfield}, the initial plasma-$\beta$ sets the initial poloidal magnetic flux. Hence, we vary the $\beta$ of the fiducial MRN-PAH simulation (MRN-PAH-b3-A) to gauge the effect magnetic flux has on the formation of rings and gaps, similar to the corresponding study in \citet{SurianoEtAl2018}. The comparison performed in this study differs slightly however, due to the inclusion of a chemical network. Figure \ref{fig:discInitBeta} displays the initial mid-plane $\ElsasserA$ for the MRN-PAH grain distribution for $\beta = 10^2$, 10$^3$ and 10$^4$. In the original study, the neutral-field coupling profile remains unchanged when varying $\beta$, as contributions from the field-strength in the Alfv\'{e}n speed ${\rm v}_{\rm A}$ and $\etaA$ cancel each other out via equation (\ref{eqn:lambdaa}). However, with the inclusion of a chemical network, $\etaA$ is now defined by equation (\ref{eqn:etaAnew}) and the dependence on $B$ is no longer simply quadratic. As a result, when including chemistry, $\Lambda_{\rm A, 0}$ is now dependent on $\beta$, as displayed in Figure \ref{fig:discInitBeta}. 

% Describing the global figure
Figure \ref{fig:discEvolBeta} displays the evolved disc and wind morphologies for each of the simulations with different plasma-$\beta$ at time $t/t_0 = 2500$. For lower $\beta$ (higher magnetic flux, left panel), the wind is faster, more streamlined, and no accretion streams are visible. The wind density is reflective of its footprint in the disc; winds launched from the ring surfaces are much more dense than those originating in the gaps. At higher $\beta$ (lower magnetic flux, right panel) the wind is slower and more turbulent, with multiple surface accretion streams disrupting the flow inwards of 30 au.

% Describing the SD figure
The normalized surface density maps of the different $\beta$ simulations are shown in Figure \ref{fig:discSDBeta}. At lower $\beta$ the rings are much larger in radius and have higher contrast, while at higher $\beta$  ring formation is suppressed outside of 12 au, and the contrast of rings within this radius is much lower. According to the reconnection-driven ring formation model of \citet{SurianoEtAl2018}, stronger poloidal fields drive faster accretion in the gaps, allowing for a more complete depletion of disc material and creating wider gaps with lower column densities. Similarly, the stronger magnetic flux in the gaps creates accretion bottlenecks, leading to increased density within the rings.

% Discussion
Compared to the stable, periodic rings seen in the $\beta = 10^3$ case, the rings in the $\beta = 10^2$ model appear to reflect a more chaotic configuration. As shown in Figure \ref{fig:discInitBeta}, the entire simulation mid-plane initially has $\ElsasserA > 1$. \citet{SurianoEtAl2018} showed that increased coupling above $\ElsasserA = 1$ results in the development of MRI `channel flows' which dominate the disc and the wind and drive both to an unsteady state. Also, we see that for the $\beta = 10^4$ simulation, rings have all but disappeared, and this corresponds to a mid-plane $\ElsasserA < 1$. This reinforces the observation from previous sections that for stable, periodic rings structures to form the effective neutral-field coupling must be $\Elsasser \gtrsim 1$.

%=================================================
% DISCUSSION
%=================================================
\section{Discussion} \label{sec:discussion}

%Introduction
In this work we examine the ability of protoplanetary discs including chemistry and different grain size distributions to form disc substructures similar to those observed in other MHD models \citep[e.g.][]{JohansenEtAl2009, DittrichEtAl2013, KunzLesur2013, BaiStone2014, SimonArmitage2014, Bai2015, BethuneEtAl2016, BethuneEtAl2017, SurianoEtAl2018}. We perform a series of numerical simulations with varying grain size distributions and diffusivity components/plasma-$\beta$ to constrain the physical disc conditions required for the formation of rings and gaps.

% Grain distribution results summary
In Section \ref{sec:discEvol}, we observe the formation of rings and gaps in protoplanetary discs for a number of different grain size distributions. Beginning with a standard MRN grain distribution between 0.005 and 1 $\mu$m, and increasing the size of included grains to a more evolved population (eMRN) between 1 and 100 $\mu$m, we find that the prevalence of rings increases. Adding a population of PAHs is also conducive to the formation of rings, as adding tiny grains switches the main charge carrier from ions and electrons to PAHs, with the effect of reducing the magnetic diffusion levels to be on par with that of the eMRN distribution.

% Ambipolar diff sims
The different grain populations used to simulate purely ambipolar-diffusive discs span a wide range of neutral-field coupling values, from $\ElsasserA = 10^{-4}$ to $10^{1}$. From our results we find that grain populations inducing neutral-field couplings of $\ElsasserA \gtrsim 1$ are the most likely to form periodic, long-lived ring structures. The direct composition of the grain population does not seem to matter so much as the resulting value of $\ElsasserA$. Models with $\ElsasserA \ll 1$ (MRN, trMRN) are characterised by rapid removal of magnetic flux and the lack of sufficient coupling between the field and the material prevents the formation of any notable structures. On the other hand, models with $\ElsasserA > 1$ are characterised by an increased prevalence of accretion streams and non-periodic rings. 

% Ohmic diff sims
In the first part of Section \ref{sec:ohmicbeta} we add Ohmic diffusion to each of the models from Section \ref{sec:discEvol} and observe the effect Ohmic diffusion has on ring formation. We find that the inclusion of Ohmic diffusion has different effects on each of the models, depending on the change in total magnetic diffusion. In fact, we find that the $\ElsasserA \gtrsim 1$ condition for ring formation found for ambipolar-only models can be extended to those including Ohmic diffusion, replacing $\ElsasserA$ with the effective neutral-field coupling of the ambipolar-Ohmic system $\Elsasser$ (see equation \ref{eqn:effectivelambda}), giving $\Elsasser \gtrsim 1$. It is also worth noting that while the addition of PAHs to the grain population is conducive to the formation of rings in the ambipolar-only regime over their non-PAH counterparts, when Ohmic diffusion is also included, the advantage for ring formation is neutralised by the larger associated Ohmic component of the effective diffusion.

% Beta sims
Finally, we examine the effect of changing the plasma-$\beta$ of the MRN-PAH-b3-A simulation on the formation of rings. We find that a smaller $\beta$ is associated with more streamlined winds, and larger and denser rings, while larger $\beta$ makes the winds more chaotic and suppresses most ring formation. This directly aligns with what we would expect from the results found in the remainder of the paper: smaller $\beta$ leads to greater magnetic flux and higher $\Elsasser$, while larger $\beta$ does the opposite (see Figure \ref{fig:discEvolBeta}).

%-------------------------------------------------
% Comparison to other works
%-------------------------------------------------
\subsection{Comparison to other works} \label{sec:dis_theory}

% Does the Elsasser > 1 condition apply in other studies?
While the $\Elsasser \gtrsim 1$ condition proposed in this paper is observed consistently in our simulations, we must be careful before applying it as a general rule. Nevertheless we can look to other studies to give credence to our result. \citet{SurianoEtAl2018} produce a number of different ambipolar-only simulations characterised by different values of initial $\Lambda_{\rm 0}$. They find ring formation occurs in all simulations for which $\Lambda_{\rm 0} \geq 0.05$, even though the initial $\Elsasser$ doesn't initially exceed unity in the inner half of the radial domain. We also see for example in our own work, the initial $\Elsasser$ profile of S18-b3-AO in Figure \ref{fig:discInitChem} stays below unity, yet still results in very weak periodic rings. \citet{HuEtAl2022} use a similar setup to \citet{SurianoEtAl2018} with $\Elsasser_{A0} = 0.25$ and \citet{RiolsEtAl2020} use $\Elsasser_{A0} = 1$, both exhibiting rings. Hence we find that the above condition may not be so strict. More research is needed to assess what creates the ideal conditions for ring growth.

% Why don't we see rings in other simulations?
As discussed by \citet{SurianoEtAl2018}, many previous simulations do not observe the formation of rings and gaps \citep[e.g.][]{BaiStone2017, Bai2017}, potentially due to the fact that a weaker initial poloidal field is used ($\beta = 10^5$), which is shown to reduce the efficiency of ring formation. In contrast, recent ring-formation studies \citep[e.g.][]{BethuneEtAl2017, SurianoEtAl2018, HuEtAl2022} with $\beta = 10^2 - 10^4$, are known for predicting accretion flows with large accretion rates ($\sim 10^{-6}-10^{-7}$ $\msun$ yr$^{-1}$), expected for the youngest, embedded protostars \citep[e.g.][]{YenEtAl2017}, while those modelling at $\beta = 10^5$ are focussing more on the regime of classical T-Tauri stars at $\mdota = 10^{-8}$ \citep{HartmannEtAl2016}. 

% How do our ring sizes compare to other studies?

% Discuss dust trapping at pressure maxima
While we cannot model it with our current simulations due to assumptions on the dust populations, we expect that dust would accumulate in the rings due to the presence of pressure maxima (see Section \ref{sec:ringstruct}). Both \citet{RiolsEtAl2020} and \citet{HuEtAl2022} observe dust accumulation in rings within their simulations when including dust transport, although \citet{HuEtAl2022} conclude that this is not due to pressure gradients but is driven primarily by dust advection from the accretion channel and subsequent meridional flows within the disc. Dust accumulation into rings may also be dependent on the dominance of mid-plane accretion, in which case the surface-dominant accretion modes seen by \citet{RiolsEtAl2020} and within this study could provide a good test to determine which mechanism is ultimately responsible.

% Discussing the ring formation mechanism
With regard to the formation mechanism generating the ring/gap structures in the disc, there are two dominant theories. \citet{SurianoEtAl2018} propose a mechanism for separating vertical magnetic flux and matter that is dependent on the reconnection of highly pinched poloidal fields around a mid-plane current sheet. However, similar to the findings of \citet{RiolsEtAl2020}, we only observe pinched poloidal fields within the first 100$t_0$ in our models including chemistry, and 
% very minimal 
relatively few clear reconnection events, which occur in both radial directions. 
%We also find that accretion flows and corresponding current sheets can occur both at the mid-plane and disc surface, in opposition to the requirements of the reconnection paradigm. 
Overall, our findings appear to have a greater agreement with the wind-driven ring formation model presented by \citet{RiolsLesur2019}, in which ring formation is a natural consequence of MHD discs launching winds.

% Discussing the addition of Ohmic diffusion and impact of PAHs
In this paper, we find that ring formation only occurs for $\Elsasser \gtrsim 1$. In the ambipolar diffusion limit, which is generally expected to dominate for $r > 10$ au \citep{KoniglSalmeron2011}\footnote{If the Hall effect becomes important, it can strongly affect, or even dominate, the poloidal magnetic transport in the disk \citep[e.g.][]{BaiStone2017}. There is some indication from global non-stratified disk simulations that the Hall effect can concentrate the poloidal magnetic flux into narrow rings by itself \citep[e.g.][]{BethuneEtAl2017}. It appears to work against but does not prevent the AD-induced flux concentration in stratified disks \citep[e.g.][]{BethuneEtAl2017}. More work is needed to clarify the role of the Hall effect on disk substructure formation.}, we find that the presence of smaller grains down to 5 nm leads to much larger diffusion rates. When sub-nanometer PAH's are included, the diffusion rates drop substantially due to the change in main charge carrier \citep{Bai2011b}. This means that the threshold for ring formation is only satisfied for evolved dust populations, or when much smaller PAHs are present. This picture changes substantially when Ohmic diffusion is included. For non-PAH models, Ohmic diffusion only dominates over ambipolar inside of 8au, minimising the effect on the effective coupling for the majority of the disc. However, the inclusion of Ohmic diffusion for PAH models is quite dramatic. With the presence of PAHs, Ohmic diffusion is orders of magnitude larger than ambipolar, drastically reducing the coupling throughout the disc and suppressing any possible ring formation. We should stress that the PAH abundance is uncertain in protoplanetary disks. It may be possible to have a PAH abundance that is large enough to reduce the ambipolar diffusivity significantly but not so large as to drastically increase the total grain surface for recombination of free elections and thus reduce the electron fraction (and increase the Ohmic diffusivity). This potentially non-monotonic behavior of the effects of PAHs on magnetic coupling deserves further exploration in the future.
%Hence it is important to understand the motion of PAHs within discs to quantify the effects they could have on ring formation.

%-------------------------------------------------
% Comparison to observations
%-------------------------------------------------
\subsubsection{Comparison to observations} \label{sec:dis_obs}

% Discuss the implications of our discovery of the limits on grain size for ring formation
In Section \ref{sec:ohmic} we show that for discs including ambipolar and Ohmic diffusion, spontaneous ring/gap formation only occurs for larger grain populations, i.e. for those grain populations with $\amin > 1\mu$m, with the presence of smaller grains ($5$ nm < $a$ < $1\mu$m) leading to a sufficiently decoupled field such that rings cannot form, while PAHs ($a = 0.5$nm), though decreasing the ambipolar diffusion, can increase the Ohmic diffusion enough to give a similar effect. Hence we expect only to see ring formation in discs with evolved grain populations.

Rings and gaps have been frequently detected in class II protostellar objects ($\sim$ 1 Myr), and more recently in the study of \citet{SeguraCoxEtAl2020}, these structures were found in the class I source IRS 63 (< 0.5 Myr) in the nearby Ophiuchus molecular cloud. Given our results, this may imply that grain growth in protoplanetary discs proceeds at a rapid pace in very young discs.

In both our study and that of \citet{SurianoEtAl2018}, ring and gap width increases with smaller plasma$-\beta$ and larger field-neutral coupling $\Elsasser$. This may give some predictive power with regards to the conditions within observed discs. For example, in \citet{SeguraCoxEtAl2020} the IRS 63 disc has two prominent rings located at 27 and 51 au, giving a radial position ratio of $\sim$ 2:1. Position ratios for other discs, such as that of \citet{SierraEtAl2021} are in the range 1.5:1 to 2:1. According to our results, these structures are best fit by disc models with higher $\Elsasser$ and more evolved grain distributions, and discs with lower plasma$-\beta$.

%-------------------------------------------------
% Limitations and future work
%-------------------------------------------------
\subsection{Future directions}

% Introduction
There are a number of additions to the physical models presented here that can be pursued following this investigation. 
% Hall effect, if the added diffusion still allows ring formation, and noting the interesting results of BethuneEtAl2018.
The first is to include the Hall effect. Given the effect of the addition of Ohmic diffusion to ring formation in Section \ref{sec:ohmic}, adding another source of diffusion could lead to full quenching of ring formation in the regions of the disc where Hall diffusion is significant. Additionally, \citet{BethuneEtAl2017} included Hall diffusion in their models and found that ring formation was suppressed when including a reversed background field (i.e. $\mathbf{\Omega} \cdot \mathbf{B} < 0$). This observation is worth a more thorough study to confirm.
% More accurate chemistry modelling, including grain migration, growth and radiative transfer to give more accurate chemistry in the gaps, which could be observed in future.
Additionally, a more detailed treatment of the thermodynamics of the disk atmosphere is desirable since it can affect the mass loss rate of the disk wind (e.g., \citealt{2019ApJ...874...90W}). This effect is expected to become more important for less magnetized disks where the thermal pressure gradient plays a bigger role in wind launching. Furthermore, a more accurate chemical network, including grain migration and coagulation, plus simple radiative transfer would increase the chemically predictive power of the model, providing chemical abundances in the magnetically dominant gaps, which could be observed in the near future. Finally, it would be interesting to extend the current 2.5D (axisymmetric) simulations to 3D, where the gas substructure is expected to have an azimuthal dependence. Azimuthal variations are indeed found in the 3D simulations of Suriano et al. (2019) using a prescribed ambipolar diffusion, but they do not grow to such an extent as to disrupt the rings and gaps. It would be important to quantify how the azimuthal variations of the gas affect the appearance of dust rings and gaps directly probed by dust continuum observations.

%=================================================
% CONCLUSIONS
%=================================================
\section{Conclusions} \label{sec:conclusions}

% Introduction
% This is basically an overview of what we were trying to achieve and what methodology we used. Perhaps use the abstract I included in the Fachbeirat?
In this paper, we build on previous work investigating ring formation in non-ideal MHD protoplanetary disks.
%protoplanetary discs via non-ideal MHD instabilities. 
We remove the parametrization of non-ideal terms present in these studies by including a simple chemical network and grain distribution model to calculate the non-ideal effects in a more self-consistent way. We simulate ring formation for different disc conditions by using a range of grain distributions, and gauge the influence of ambipolar and Ohmic diffusion components and changes to the mid-plane plasma-$\beta$, on ring formation in the disc. We uncover a general condition for ring formation based on the neutral-field coupling $\Elsasser$, and which grain populations are likely to reach this threshold and trigger ring growth. We also discuss the implications for this in detail. In summary, we find the following:

% Dot points of main discoveries
\begin{itemize}
\item[(i)] We find that a general requirement of $\Elsasser \gtrsim 1$ is needed for stable ring and gap formation within PPDs. This requirement is observed when varying grain distribution, with the implementation of different non-ideal diffusion regimes and when modifying the plasma-$\beta$. This condition applies to the initial conditions of the disc mid-plane and must be satisfied for the majority of disc radii for ring formation to occur. Discs for which this condition is not satisfied have insufficient coupling between the magnetic field and accreting material, reducing the likelihood that the field will radially compress enough to trigger ring formation. For discs with $\Elsasser \ll 1$ the coupling is so low that the Lorentz force is much greater than the drag force from the disc material, allowing the field to rapidly move outwards. Discs with $\Elsasser \gg 1$ form rings, however these rings have much greater contrast than those formed at $\Elsasser \gtrsim 1$ and are less periodic due to the chaotic nature of the MRI, which is triggered in this regime.
\item[(ii)] Matching certain grain populations to their corresponding $\Elsasser$ profiles in a toy disc including both ambipolar and Ohmic diffusion, we find that for our chosen disc parameters, only evolved grain populations with micrometer and larger sized grains are able to produce the necessary coupling to allow the formation of rings within discs.
\item[(iii)] The inclusion of PAHs in ambipolar diffusion-dominated discs enhances the ability of discs with submillimeter grains to form rings by changing the dominant charge carriers from ions and electrons to grains. This dramatically increases $\Elsasser$ and moves the disc into the ring-forming regime. In contrast, when Ohmic diffusion is included, PAHs have the opposite affect, suppressing almost all ring formation.
\end{itemize}

% Conclusions
% Broad conclusions with a reminder to look to more accurate modelling in the future.
In conclusion, we find that the chemical and dust composition of the disc is particularly important within the context of ring formation in protoplanetary discs, and as a consequence, in whether young systems can form planets. We do stress though that more research is necessary to confirm the findings of this paper, for example by including Hall diffusion and a more thorough treatment of the chemistry.

%%%%%%%%%%%%%%%%% ACKNOWLEDGEMENTS %%%%%%%%%%%%%%%
\section*{Acknowledgements}
We thank Scott Suriano for the use of his initial simulation configurations.
We also thank Christian Rab, Geoffroy Lesur and Kees Dullemond for useful discussions and the referee for a detailed and constructive report. CN and PC acknowledge the financial support of the Max Planck Society. ZYL is supported in part by NASA 80NSSC20K0533 and NSF AST-2307199 and AST-1910106.

%%%%%%%%%%%%%%%%%%%%%%%%%%%%%%%%%%%%%%%%%%%%%%%%%%
\section*{Data Availability}
The data underlying this article will be shared on reasonable request to the corresponding author.



\bibliographystyle{style/mnras}
\bibliography{references/references.bib} % if your bibtex file is called example.bib

%\begin{thebibliography}{99}
%\end{thebibliography}

%%%%%%%%%%%%%%%%%%%%%%%%%%%%%%%%%%%%%%%%%%%%%%%%%%

%%%%%%%%%%%%%%%%% APPENDICES %%%%%%%%%%%%%%%%%%%%%

\appendix

%-------------------------------------------------
% Derivation of Elsasser number for multiple diffusivity regimes
%-------------------------------------------------
\section{Derivation of \texorpdfstring{$\Elsasser$}{Lambda}} \label{sec:devdiff}

Here we briefly derive the effective Elsasser number $\Elsasser$ as a function of the Ohmic, Hall and ambipolar Elsasser numbers. The effective Elsasser number, the ratio of the Lorentz to Coriolis forces, which measures the degree of coupling between the neutrals and the magnetic field is defined by
\begin{equation}
\Elsasser = \frac{{\rm v}_{\rm A}^2}{\Omega_{\rm K} \eta_{\perp}} \label{eqn:elsassfull}
\end{equation}
\citep{KoniglEtAl2010}, where 
\begin{equation}
\eta_{\perp} = \frac{c^2}{4 \pi \sigma_{\perp}}
\end{equation}
is the perpendicular magnetic diffusivity, based on the total conductivity perpendicular to the magnetic field
\begin{equation}
\sigma_{\perp} = \sqrt{\sigma_{\rm H}^2 + \sigma_{\rm P}^2}.
\end{equation}
Introducing the definitions of the Hall and Pederson diffusion coefficients 
\begin{eqnarray}
\etaH = \frac{c^2}{4 \pi} \frac{\sigma_{\rm H}}{\sigma_{\perp}^2}; \qquad \eta_{\rm P} = \etaO + \etaA = \frac{c^2}{4 \pi} \frac{\sigma_{\rm P}}{\sigma_{\perp}^2}, \label{eqn:etahp}
\end{eqnarray}
and associated Elsasser numbers
\begin{eqnarray}
\ElsasserH = \frac{{\rm v}_{\rm A}^2}{\Omega_{\rm K} \etaH}; \qquad \ElsasserP = \frac{{\rm v}_{\rm A}^2}{\Omega_{\rm K} \eta_{\rm P}}, \label{eqn:elsasshp}
\end{eqnarray}
we can rearrange them to give an alternate definition for the perpendicular magnetic diffusivity:
\begin{equation}
\eta_{\perp} = \sqrt{\eta_{\rm P}^2 + \etaH^2}. \label{eqn:etaperp}
\end{equation}
For completeness, the ambipolar and Ohmic diffusion coefficients are
\begin{eqnarray}
\etaO = \frac{c^2}{4 \pi \sigma_{\rm O}}; \qquad \etaA = \frac{c^2}{4 \pi \sigma_{\perp}} \frac{\sigma_{\rm P}}{\sigma_{\perp}} - \etaO,
\end{eqnarray}
with associated Elsasser numbers
\begin{eqnarray}
\ElsasserO = \frac{{\rm v}_{\rm A}^2}{\Omega_{\rm K} \etaO}; \qquad \ElsasserA = \frac{{\rm v}_{\rm A}^2}{\Omega_{\rm K} \etaA}.
\end{eqnarray}
It follows that
\begin{equation}
\frac{1}{\ElsasserP^2} = \left(\frac{1}{\ElsasserO} + \frac{1}{\ElsasserA}\right)^2 \label{eqn:elsasspao}
\end{equation}
Substituting equation (\ref{eqn:etaperp}) into (\ref{eqn:elsassfull}) and employing equations (\ref{eqn:elsasshp}) and (\ref{eqn:elsasspao}) we arrive at
\begin{equation}
\Elsasser = \left(\frac{1}{\ElsasserH^2} + \left(\frac{1}{\ElsasserO} + \frac{1}{\ElsasserA}\right)^2\right)^{-1/2}. \label{eqn:elsassfull2}
\end{equation}
In the regime including only ambipolar and Ohmic diffusion ($\ElsasserH = \infty$), the effective Elsasser number reduces to
\begin{equation}
\Elsasser = \frac{\ElsasserO\ElsasserA}{\ElsasserO + \ElsasserA}. \label{eqn:elsassoa}
\end{equation}

%%%%%%%%%%%%%%%%%%%%%%%%%%%%%%%%%%%%%%%%%%%%%%%%%%

% Don't change these lines
\bsp	% typesetting comment
\label{lastpage}
\end{document}